\documentclass{article}

{}
{}
{}
\usepackage{amssymb,amsmath}
\usepackage[ruled,vlined]{algorithm2e}
\usepackage{array}
\usepackage{bm}
\usepackage{xcolor}
\usepackage{graphicx}
\usepackage{caption}
\usepackage{subcaption}
\usepackage{url,soul}

\begin{document}

%\supertitle{Submission Template for IET Research Journal Papers}
\begin{titlepage}
	\vspace*{\stretch{1}}
	\begin{center}
		{\Large\bfseries Space Alternating Variational Estimation Based Sparse Bayesian Learning for Complex-value Sparse Signal Recovery Using Adaptive Laplace Priors}                  \\[6.5ex]
		{\large\bfseries Zonglong Bai$^{1,2,3}$,  Liming Shi$^{4,5}$, Jinwei Sun$^{3}$, Mads Gr\ae sb\o ll Christensen$^{5}$}           \\
		\vspace{4ex}
		1. Department of Electronic and Communication Engineering, North China Electric Power University\\
		2. Hebei Key Laboratory of Power Internet of Things Technology, North China Electric Power University\\
		3. School of Instrument Science and Engineering, Harbin Institute of Technology\\
		4. Department of Communication and Information Engineering, Chongqing University of Posts and Telecommunications\\
		5. CREATE, Aalborg University\\[5pt]
		\textit{Postal and e-mail address of Zonglong Bai: \\
			Baoding 071003, Hebei, China\\
		baizongyao@163.com}                \\
		\textbf{Conflict of interest statement}: the authors declare that they have no competing interest.\\
		\textbf{Funding information}: this work was supported in part by the Fundamental Research Funds for the Central Universities (Grant No. 2022MS077).\\
		\textbf{Data availability statement}: data available on request from the authors. \\
		\textbf{Credit contribution statement}: Z. Bai and L. Shi conceptualized the study and run the experiments; M. G. Christensen and J. Sun edited the manuscript; All the authors read and approved the final manuscript.
		\vfill
		\today
	\end{center}
	\vspace{\stretch{2}}
\end{titlepage}

\title{}
\date{}
%\author{1. Department of Electronic and Communication Engineering, North China Electric Power University, Baoding 071003, Hebei, China\\
%	2. Hebei Key Laboratory of Power Internet of Things Technology, North China Electric Power University, Baoding 071003, Hebei, China\\
%	3. School of Instrument Science and Engineering, Harbin Institute of Technology,Harbin 150001, Heilongjiang, China\\
%	4. Department of Communication and Information Engineering, Chongqing University of Posts and Telecommunications, Chongqing 400065, Chongqing, China\\
%	5. Department of Architecture, Design and Media Technology, Aalborg University, Aalborg 9000, Aalborg, Denmark}
\maketitle
%\add{2}{Second Company Department, Company Address, City, Country Name}
%\add{3}{Third Department, Third University, Address, Country Name}
%\add{4}{Current affiliation: Fourth Department, Fourth University, Address, Country Name}
%\email{baizongyao@163.com}}

\begin{abstract}
Due to its self-regularizing nature and its ability to quantify uncertainty, the Bayesian approach has achieved excellent recovery performance across a wide range of sparse signal recovery applications. However, most existing methods are based on the real-value signal model, with the complex-value signal model rarely considered. Motivated by the adaptive least absolute shrinkage and selection operator (LASSO) and the sparse Bayesian learning (SBL) framework, a hierarchical model with adaptive Laplace priors is proposed in this paper for recovery of complex sparse signals. Moreover, the space alternating approach is integrated into the algorithm to reduce the computational complexity of the proposed method. In experiments, the proposed algorithm is studied for complex Gaussian random dictionaries and different types of complex signals. These experiments show that the proposed algorithm offers better recovery performance for different types of complex signals than state-of-the-art methods.
\end{abstract}

\maketitle

\section{Introduction}

Sparse signal recovery (SSR) aims to recover the sparse signal from a set of linear measurements. It is a fundamental problem in signal processing with practical utility in a wide range of applications. Some applications, such as image processing and electroencephalography (EEG) \cite{JIANG2020101966,ZHU2021209,8695874,ABIANTUN2019308}, are based on real-value measurements. In other applications, such as radar signal processing, spectral estimation, blind source separation and magnetic resonance imaging \cite{9716156,9729551,8701504,8962388,Yang_2018}, the complex-value signal model is often used, as the phase information is then readily available and can be separated from the amplitude. In such settings, the complex-value model can be seen as an extension of the real-value model, relying on the prior knowledge that the real part and the image part are jointly sparse. Although a complex-value model can be converted into a real-value model, and the real and image parts can be recovered separately \cite{Carlin2013,Bai2019}, the joint sparsity of the real and image parts should be considered in the recovery process. Moreover, the dimensions of the dictionary and measurements are effectively doubled when converting the complex model into a real one, resulting in a higher computational complexity. To deal with this problem, we propose a computational efficient complex-value sparse signal recovery algorithm in this paper.

The least absolute shrinkage and selection operator (LASSO) and its variants have been widely applied to the sparse signal recovery problem \cite{Tibshirani1996,Maleki2013,Zou2006,Candes2008a,Tibshirani2011}. In \cite{Tibshirani1996}, the LASSO is proposed for estimation in linear models. The $l_1$ norm is used as the penalty, which limits the the solution space to a finite subset. Considering the complex-value model, a complex LASSO algorithm is proposed in \cite{Maleki2013}, where the approximate message passing approach is used to enforce the prior knowledge of the joint sparsity. However, the LASSO is not an oracle procedure \cite{Zou2006}. To enforce the oracle properties, as defined in \cite{Fan2001}, an adaptive LASSO is proposed in \cite{Zou2006}. Instead of using a common regularized factor, a series of data-dependent weights are assigned to different coefficients. In signal processing, the approach of minimizing the $l_1$ norm penalized problem is often referred to as basis pursuit. Besides these $l_1$ norm optimization methods, greedy algorithms are also often used for sparse signal recovery. These include examples such as orthogonal matching pursuit \cite{Tropp2007,Donoho2012},  newtonized orthogonal matching pursuit (NOMP) \cite{7491265,9181410}, greedy pursuit \cite{Blumensath2008}, cyclic matching pursuit \cite{Christensen2007}, and subspace pursuit \cite{Dai2009}. 

Both the LASSO methods and the greedy algorithms are essentially deterministic regularization approaches that provide only a point estimate. The Bayesian framework can also be used to formulate the sparse signal recovery problem. This framework is a probabilistic prediction approach that provides both the estimation of model parameters and the estimation of the associated uncertainty. Additionally, all of the model parameters can be updated automatically when using this method, due to its self-regularizing nature. From a Bayesian perspective, the LASSO is equivalent to building a Bayesian model with Laplace priors \cite{Tibshirani1996}. The underlying assumption is that all of the parameters are Laplace-distributed with a common scale factor. However, when considering the conjugate prior principle, the Laplace distribution is not a conjugate prior of the Gaussian distribution. In other words, the problem of determining the posterior distribution is intractable if Laplace priors are used directly. In \cite{Tipping_2001}, a hierarchical Bayesian framework is proposed, called sparse Bayesian learning (SBL). In this hierarchical Bayesian framework, each element of the unknown signal is assigned an independent zero-mean Gaussian distribution with precision as the hidden parameter. Furthermore, a gamma distribution is imposed on the precision parameter in the second layer. This particular hierarchical model is equivalent to adding student-t priors to the likelihood. The student-t prior has similar properties as the Laplace prior, but ensures that the the hidden parameter distributions is a conjugate prior. The student-t prior is thus convenient for calculating the posterior distribution of each hidden parameter. The hierarchical Bayesian framework is further applied into compress sensing in \cite{Ji2008}.   In \cite{Babacan2010}, a hierarchical model of Laplace priors is proposed. In the first layer, a zero-mean multivariate Gaussian distribution is used to describe the unknown signals, following the SBL framework. In the second stage, independent inverse Gamma priors are assigned to the precision of the Gaussian distributions. The marginal distributions with respect to the precisions that results from the two stages in the hierarchy lead to Laplace priors. Furthermore, a common Gamma prior is assigned to all of the hyper-parameters of the second-stage priors. This hierarchical framework can be seen as a Bayesian perspective on LASSO. In \cite{Themelis2012} and \cite{Leng2014}, the hierarchical model is further improved. Independent Gamma priors are applied to the hyper-parameters of the second-stage hierarchy, which results in a new hierarchical Bayesian model corresponding to the real-value adaptive LASSO. Then, a variational Bayesian sparse signal recovery method with Laplacian scale mixture prior is proposed in \cite{Zhang2017a}. For the complex-value case, an autofocus algorithm using Laplace priors is proposed in \cite{Zhao2014} for synthetic aperture radar imagery. In our previous work, we studied acoustic direction of arrival (DOA) estimation using SBL with student-t prior to get high resolution performance \cite{EURASIP,WASPAA,EUSIPCO}. 

In the aforementioned Bayesian approaches, a matrix inverse operation is required, something that leads to a high computational complexity. To overcome this problem, a basis addition and deletion approach is proposed to accelerate the evidence maximization (type-II maximum likelihood) procedure, which results in a faster algorithm \cite{Tipping03fastmarginal}. In \cite{Babacan2010}, the criterion of basis addition and deletion was improved for the Laplace priors based Bayesian model. In \cite{Zheng2015}, the space alternative method is used to accelerate a variational Bayesian inference approach. In \cite{Thomas2018}, the space alternative-based method is further improved resulting in the so-named space alternative variational estimation (SAVE) method. Furthermore, a scalable mean-field SBL technology is proposed in \cite{Worley2019} for solving large size problems based on the space alternative approach. In \cite{8410591}, an SBL algorithm is proposed for massive multiple-input multiple-output (MIMO) channel estimation and the Kalman filter (KF) and Rauch–Tung–Striebel smoother (RTSS) are used to accelerate the SBL algorithm. In \cite{9110823}, a fast SBL algorithm is proposed and a basis addition and deletion strategy is used to reduce the computational complexity.
 
It is worth mentioning that most of the aforementioned methods are based on the real-value signal model. To the best of the authors knowledge, no close form criterion for the complex-value signal model has been reported or proposed in the literature. The complex-value signal model is rarely considered in neither Bayesian methods nor some of the methods based on convex optimization\footnote{For example, basis pursuit is based on linear programming, but the complex problem cannot be cast as a linear programming problem. Thus, it cannot be directly applied the complex-value signal model. }. We argue that it is useful to consider the complex problem because the complex-value model is widely used in practice. In this paper, we build a hierarchical Bayesian framework for the complex-valued signal model. 
The contributions of this paper are listed as follows:
\begin{itemize}
	\item Inspired by the hierarchical model with Laplace priors proposed in \cite{Babacan2010} and the adaptive LASSO proposed in \cite{Zou2006}, we develop a hierarchical Bayesian model with adaptive Laplace priors for complex-value signal model, which can be used to improve the recovery accuracy performance and DOA estimation performance.
	\item  To avoid the matrix inverse operation, the space alternative method is integrated into the proposed method, thereby reducing the computational complexity of the algorithm.
\end{itemize}

The rest of this paper is organized as follows. In Section \ref{sec:background}, the background and problem formulation of sparse signal recovery are given. In Section \ref{sec:BayesianModel}, we first give the hierarchical framework of complex Laplace priors. Then, the hierarchical model using adaptive Laplace priors is proposed for complex-value signal. In section \ref{sec:VBI}, variational Bayesian inference is used to update the hidden parameters in the proposed hierarchical Bayesian model, and the space alternation method is integrated into the proposed algorithm to avoid matrix inverse operations. In section \ref{sec:exp}, the performance of the proposed method is tested using complex Gaussian random dictionaries for complex Gaussian signals, complex Laplace signals and complex spike signals, respectively. Moreover, we apply the proposed algorithm to acoustic DOA estimation. The conclusion is given in  section \ref{sec:conclusion}. 

Throughout this paper, the bold symbols in lowercase and uppercase font are reserved for vectors and matrixes, respectively. $ \|\cdot\|_p$ and $ \|\cdot\|_f$ denote the $l_p$ norm and matrix Frobenius norm, respectively. $(\cdot)^{\rm H}$ and $(\cdot)^{*}$ denote the conjugate transpose operation and conjugate operation, respectively. $\odot$ denotes the element product operator. ${\rm diag}(\bm{v})$ denotes a diagonal matrix with a given vector $\bm{v}$ as the diagonal elements. $\mathcal{CN}(\bm{x};\bm{\mu},\bm{\Lambda})$ denotes that the variable $\bm{x}$ follows a multivariate normal distribution with the mean $\bm{\mu}$ and the variance $\bm{\Lambda}$. $\mathcal{G}(\bm{x};a,b)$ denotes that the variable $\bm{x}$ follows a Gamma distribution with the shape parameter $a$ and the rate parameter $b$. $\Gamma(a)$ denotes a value from the Gamma function. ${\rm E}_{q(\bm{\theta})}(\cdot)$ denotes the expectation with the distribution $q(\bm{\theta})$. $\mathbb{C}^M$ and $\mathbb{C}^{M\times N}$ denote the set of $M$-dimensional complex vectors and the set of complex matrixes with $M$ rows and $N$ columns, respectively. 

\section{Background and problem formulation}\label{sec:background}
In this section, we will provide some background in the form of signal models and then arrive at a problem formulation of the sparse recovery problem. In what follows, we consider both a single-measurement vector (SMV) signal model and a multiple-measurement vectors (MMV) signal model. For the SMV case, the prior knowledge entails the fact that a few elements of the unknown signals are non-zero while the others are zero. For the MMV case, we assume that a few rows of the unknown signals are non-zero, i.e., the unknown signals possess group sparsity for the MMV case.

\subsection{Signal model}
\subsubsection{Signal model for the SMV case}
Consider the problem of recovering a sparse signal $\bm{g}\in\mathbb{C}^N$ from a set of noisy under-sampled linear measurements $\bm{x}\in\mathbb{C}^M$ with the observation model as follows: 
\begin{equation}
\bm{x}=\bm{A}\bm{g}+\bm{w},\label{sigModel}
\end{equation}
where 
\begin{align}
&\bm{x}= [x_1,\cdots,x_m,\cdots,x_M]^{\rm T},\nonumber\\
&\bm{g}= [g_1,\cdots,g_n,\cdots,g_N]^{\rm T},\nonumber\\
&\bm{w}= [w_1,\cdots,w_m,\cdots,w_M]^{\rm T},\nonumber
\end{align}
and $m$ is the index of the measurement elements, $M$ is the total number of elements in $\bm{x}$, i.e., the length of measurement. Moreover, $n$ is the index of the unknown signal elements, $N$ is the length of the unknown signals,  $\bm{A}\in\mathbb{C}^{M\times N}$ denotes the dictionary ($M\leqslant N$), and $\bm{w}\in\mathbb{C}^M$ denotes the noise. Given the measurement $\bm{x}$ and the dictionary $\bm{A}$, we try to recover the unknown source signal $\bm{g}$ as accurately as possible. 

\subsubsection{Signal model for the MMV case}
Let $\bm{X}\in\mathbb{C}^{M\times L}$ represents the measurements and $\bm{G}\in\mathbb{C}^{N\times L}$ denotes the unknown signal for the MMV case. Then, the observation model for the MMV case can be described as 
\begin{equation}
\bm{X}=\bm{A}\bm{G}+\bm{W},
\end{equation}
where
\begin{align}
&\bm{X}= \left[\bm{x}_{\cdot 1},\cdots,\bm{x}_{\cdot l},\cdots,\bm{x}_{\cdot L}\right],
&\bm{x}_{\cdot l}= \left[x_{1l},\cdots,x_{Ml}\right]^{\rm T},\nonumber\\
&\bm{G}= \left[\bm{g}_{\cdot 1},\cdots,\bm{g}_{\cdot l},\cdots,\bm{g}_{\cdot L}\right] ,
&\bm{g}_{\cdot l}= \left[g_{1l},\cdots,g_{Nl}\right]^{\rm T},\nonumber\\
&\bm{W}= \left[\bm{w}_{\cdot 1},\cdots,\bm{w}_{\cdot l},\cdots,\bm{w}_{\cdot L}\right] ,
&\bm{w}_{\cdot l}= \left[w_{1l},\cdots,w_{Ml}\right]^{\rm T},\nonumber
\end{align}
$L$ is the total number of measurement vectors, and $l$ is the index of measurement vector.

\subsection{Problem formulation}
We will here consider the under-sampled measurements case where the length of measurement $M$ is smaller than the length of unknown source signals $N$. In this case, the recovery of either $\bm{g}$ or $\bm{G}$ using, e.g., ordinary least square (OLS) method is an ill-posed problem. To overcome this problem, the minimum power constraint term can be assigned to the OLS, which results in the ridge regression, as follows: 
\begin{equation}
\tilde{\bm{g}}=\arg \min\limits_{\bm{g}}\left\|\bm{x}-\bm{A}\bm{g}\right\|_2^2+\eta\left\|\bm{g}\right\|_2,\label{l2Norm}
\end{equation}
where $\tilde{\bm{g}}$ denotes the estimation of the unknown signal $\bm{g}$ and $\eta$ is a regularized factor. Ridge regression offers a number of computational advantages, although it spreads its energy across all entries instead of a subset. In other words, the ridge regression does not provide a sparse solution. 

To exploit the sparse prior knowledge of the unknown signal, a regularized constraint $\|\bm{g}\|_0$ is added to OLS, where the $\|\bm{g}\|_0$ norm denotes the number of non-zero elements in $\bm{g}$. This sparse constraint limits the solution space to a finite subset, which results in a more accurate recovery performance than the OLS method. As a result, the sparse signal recovery problem can be described as a minimization optimization problem, as follows:
\begin{equation}
\tilde{\bm{g}}=\arg \min\limits_{\bm{g}}\; \left\|\bm{x}-\bm{A}\bm{g}\right\|_2^2+\eta\left\|\bm{g}\right\|_0,\label{norm0Form}
\end{equation}
where $\eta$ is a predefined regularized factor. However, this optimization problem is intractable because it is a non-deterministic polynomial hard (NP-hard) problem. The most common way to manage the $l_0$-norm penalized problem is by relaxing it to the $l_1$-norm penalized problem, which results in the following minimization problem: 
\begin{equation}
\tilde{\bm{g}}=\arg \min\limits_{\bm{g}}\; \left\|\bm{x}-\bm{A}\bm{g}\right\|_2^2+\eta\left\|\bm{g}\right\|_1.\label{norm1Form}
\end{equation}
This estimator can be viewed as an $l_1$-norm penalized least square estimator, which is also known as a least absolute shrinkage and selection operator, or LASSO \cite{Tibshirani1996}. Furthermore, using a series of data-dependent weights $\bm{\varsigma}$ instead of a common weight $\eta$, the adaptive LASSO is given as \cite{Zou2006}
\begin{equation}
\tilde{\bm{g}}=\arg\min\limits_{\bm{g}}\;\left\|\bm{x}-\bm{A}\bm{g}\right\|_2^2+\eta\left\|\bm{D}\bm{g}\right\|_1,\label{ReWe_cLASSO}
\end{equation}
where $\bm{D}={\rm diag}(\bm{\varsigma})$ is the diagonal weight matrix. Because the adaptive LASSO enjoys the oracle properties\footnote{The oracle properties include the consistency of model selection and the asymptotic normality of parameter estimation. The consistency of model selection is that the correct model is selected with probability 1 when some parameters are unknown. The asymptotic normality of parameter estimation is that the estimation of the corresponding non-zero coefficient has the same optimal convergence rate as the least squares estimator under the real model. }, it provides a better basis selection performance than LASSO and thus results in better recovery performance\cite{Zou2006}.

For the MMV case, the prior knowledge includes the fact that the unknown signal has group sparsity. Similar to the SMV case, the adaptive LASSO for the MMV case is 
\begin{equation}
\tilde{\bm{G}}=\arg\min\limits_{\bm{G}}\;\left\|\bm{X}-\bm{A}\bm{G}\right\|_2^2+\eta\sum_{i=1}^{N}\varsigma_i\left\|\bm{g}_{i\cdot}\right\|_1,\label{ReWe_cLASSO_MMV}
\end{equation}
where $\bm{g}_{i\cdot}$ denotes the $i$th row of $\bm{G}$ and $\tilde{\bm{G}}$ is the estimate of $\bm{G}$.

The Bayesian representations of (\ref{norm1Form}) and  (\ref{ReWe_cLASSO}) are given in \cite{Babacan2010} and \cite{Themelis2012}, respectively. Both representations are based on the real-value signal model. Because there are several differences between the Bayesian frameworks used for the real-value and complex-value signal models, we next propose the hierarchical Bayesian models of (\ref{ReWe_cLASSO})  and  (\ref{ReWe_cLASSO_MMV}) for the complex-value signal model.

\section{Bayesian Modeling}\label{sec:BayesianModel}
In Bayesian modeling, all of the unknown variables are treated as stochastic variables. These variables are each assigned different priors which indicate the prior knowledge. Motivated by the SBL framework and the hierarchical model of Laplace priors, we propose a hierarchical Bayesian model using complex adaptive Laplace priors for complex-value signal model which is then used for both the SMV and MMV cases.   

\subsection{Bayesian model for the SMV case}\label{sec:HM-SMV}
\subsubsection{Noise model}
For the SMV case, we assume that the noise $\bm{w}$ is independent complex Gaussian noise with the following distribution: 
\begin{equation}
p(\bm{w}|\rho) =\mathcal{CN}(\bm{w};\bm{0},\rho^{-1}\textbf{I}_M),\label{noisePrior1}
\end{equation}
where $\rho$ is the precision of the noise, while $\textbf{I}_M$ denotes an identity matrix with size $M$. Based on the observation model (\ref{sigModel}) and the noise model (\ref{noisePrior1}), the likelihood can be described as
\begin{equation}
p(\bm{x}|\bm{g},\rho)=\mathcal{CN}(\bm{x};\bm{A}\bm{g},\rho^{-1}\textbf{I}_M).\label{obserPrio}
\end{equation}
For tractability, a Gamma prior is employed for the noise precision $\rho$ based on the conjugate prior principle, i.e.,
\begin{equation}
p(\rho|a,b)=\mathcal{G}(\rho;a,b),\label{noisePrior2}
\end{equation}
where $a\geq 0$ is the shape parameter, and $b\geq 0$ is the scale parameter. The mean and variance of $\rho$ are given as $\frac{a}{b}$ and $\frac{a}{b^2}$, respectively. 

\subsubsection{Signal model}
To formulate the Bayesian model with Laplace priors, we build a hierarchical  Bayesian framework for the unknown signals. In accordance with the SBL framework, the unknown signal $\bm{g}$ is assumed to follow a zero-mean multivariate complex Gaussian distribution in the first layer, i.e.,  
\begin{equation}
p(\bm{g}|\bm{\lambda})=\mathcal{CN}(\bm{g};\bm{0},\bm{\Lambda})= \prod\limits_{i=1}^N\mathcal{CN}(g_i;0,\lambda_i),\label{sigPrior1}
\end{equation}
where $\bm{\lambda}=[\lambda_1,\cdots,\lambda_i,\cdots,\lambda_N]^\text{T}$ is a variance vector and $\bm{\Lambda}$ is a diagonal matrix with the variables in $\bm{\lambda}$ as the diagonal elements. 

For the second stage in the hierarchy, we assume that the variables in $\bm{\lambda}$ follow independent Gamma distributions, i.e.,
\begin{equation}
p(\bm{\lambda}|\gamma)=\mathcal{G}(\bm{\lambda};\frac{3}{2},\frac{\gamma}{4})
=\prod\limits_{i=1}^N \mathcal{G}(\lambda_i;\frac{3}{2},\frac{\gamma}{4}).\label{sigPrior2}
\end{equation}

\begin{figure}[!t]
\centering
\includegraphics[width=2.5in]{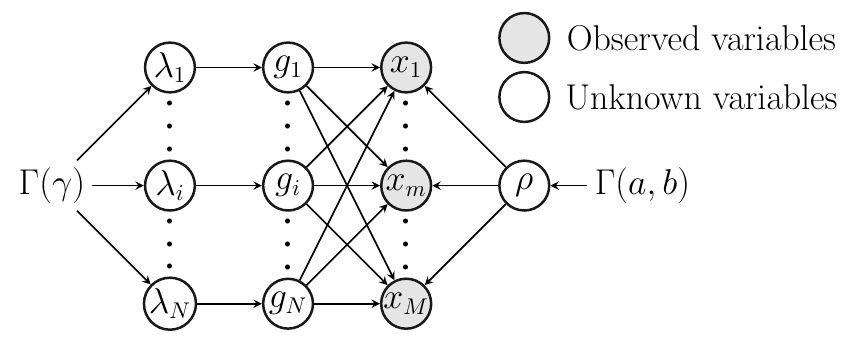}
\caption{Directed acyclic graph representation for the hierarchical model with complex Laplace priors (CL-HBM).}
\label{fig_DAG_Lap}
\end{figure}

Then, the variable $\gamma$ is assumed to follow the Gamma distribution according to the conjugate prior principle, i.e.,
\begin{equation}
p(\gamma|c,d)=\mathcal{G}(\gamma;c,d).\label{sigPrior3}
\end{equation}
Considering the first two stages in the hierarchy, we have
\begin{equation}
p(\bm{g}|\gamma)= \int p(\bm{g}|\bm{\lambda})p(\bm{\lambda}|\gamma){\rm d}\bm{\lambda}= \frac{{\gamma}^N}{(2\pi)^{N}}e^{-\sqrt{\gamma}\sum\limits_{i=1}^N|g_i|},\label{LapPrior}
\end{equation}
which is a special case of the complex generalized Gaussian distribution in \cite{Novey2010}. In this paper, we treat this case as the complex form of the Laplace prior. Assigning the prior (\ref{LapPrior}) to the likelihood (\ref{obserPrio}), the maximum a posterior (MAP) estimation is equivalent to the $l_1$-norm constraint in (\ref{norm1Form}) with the relationship $\eta=\frac{\sqrt{\gamma}}{\rho}$. The directed acyclic graph of this model is illustrated in Figure \ref{fig_DAG_Lap}.

The hierarchical model formed by (\ref{obserPrio}), (\ref{noisePrior2}), (\ref{sigPrior1}), (\ref{sigPrior2}) and (\ref{sigPrior3}) corresponds to the complex LASSO. We refer to this model as the Complex Laplace priors-based Hierarchical Bayesian Model (CL-HBM). Similarly to the LASSO, the sparsity of the unknown signal $\bm{g}$ is controlled using a common hyper-parameter $\gamma$. Thus, this model cannot meet the oracle properties. To overcome this problem, the hierarchical model is improved according to the adaptive LASSO. For the adaptive LASSO, the oracle properties are established by utilizing the adaptively re-weighted $l_1$-norm penalty. Therefore, the second stage in the hierarchy can be improved as follows:
\begin{equation}
p(\bm{\lambda}|\bm{\gamma})=\mathcal{G}(\bm{\lambda};\frac{3}{2},\frac{\bm{\gamma}}{4})= \prod\limits_{i=1}^N \mathcal{G}(\lambda_i;\frac{3}{2},\frac{\gamma_i}{4}),\label{sigPriorReWe2}
\end{equation}
where $\bm{\gamma}=[\gamma_1,\cdots,\gamma_i,\cdots,\gamma_N]^\text{T}$. 
For the third layer,  we assume that the variables in $\bm{\gamma}$ follow independent Gamma distributions, i.e.,
\begin{equation}
p(\bm{\gamma}|c,d)=\mathcal{G}(\bm{\gamma};c,d)=\prod\limits_{i=1}^N\mathcal{G}(\gamma_i;c,d).\label{sigPriorReWe3}
\end{equation}

\begin{figure}[!t]
\centering
\includegraphics[width=2.5in]{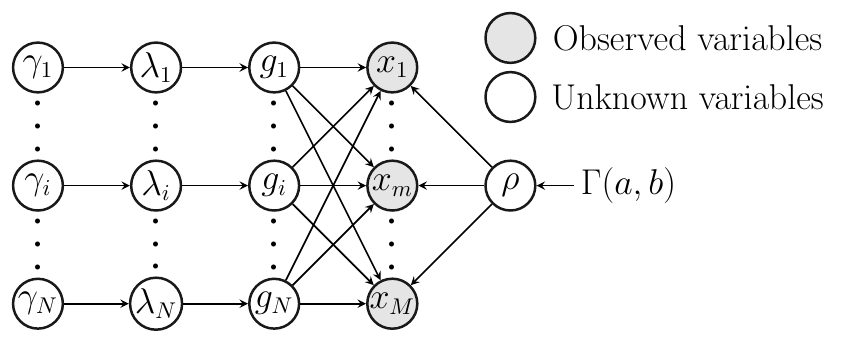}
\caption{Directed acyclic graph representation for the hierarchical model with adaptive complex Laplace priors (CAL-HBM).}
\label{fig_DAG_ReWeLap}
\end{figure}

Given the first layer (\ref{sigPrior1}) and the second layer (\ref{sigPriorReWe2}), the marginal distribution $p(\bm{g}|\bm{\gamma})$ can be obtained by marginalizing the distribution with respect to the parameter $\bm{\lambda}$ is then given by
\begin{equation}
p(\bm{g}|\bm{\gamma})= \int p(\bm{g}|\bm{\lambda})p(\bm{\lambda}|\bm{\gamma}){\rm d}\bm{\lambda}	= \frac{\prod\limits_{i=1}^N\gamma_i}{(2\pi)^{N}}e^{-\sum\limits_{i=1}^N\sqrt{\gamma_i}|g_i|}.\label{LapPriorReWe}
\end{equation}
As a result, the hierarchical Bayesian model resulting from (\ref{obserPrio}), (\ref{noisePrior2}), (\ref{sigPrior1}), (\ref{sigPriorReWe2}) and (\ref{sigPriorReWe3}) corresponds to the adaptive LASSO (\ref{ReWe_cLASSO}) with the relationship $\bm{D}={\rm diag}(\rho^{-1}\odot\sqrt{\bm{\gamma}})$, where $\sqrt{\bm{\gamma}}=[\sqrt{\gamma_1},\cdots,\sqrt{\gamma_N}]$. Therefore, data-dependent weights are assigned to each element of the unknown signal, observing the oracle properties. In this paper, we call this hierarchical model the Complex Adaptive Laplace prior-based Hierarchical Bayesian Model (CAL-HBM). Figure \ref{fig_DAG_ReWeLap} shows the hierarchical framework of the proposed CAL-HBM.

\subsection{Bayesian model for the MMV case}
As discussed in section \ref{sec:HM-SMV}, CAL-HBM is an improvement on CL-HBM. Therefore, we only consider CAL-HBM for MMV case in this part to avoid a duplicate description.
\subsubsection{Noise model}
For the MMV case, we assume that the noise is complex Gaussian noise. Then, we have
\begin{eqnarray}
p(\bm{W}|\rho)=\prod\limits_{l=1}^L p(\bm{w}_{\cdot l}|\rho)=\prod\limits_{l=1}^L\mathcal{CN}(\bm{w}_{\cdot l};\bm{0},\rho^{-1}\textbf{I}_M),\label{noisePrior1_MMV}
\end{eqnarray}
where $\bm{w}_{\cdot l}$ is the $l$'th column of $\bm{W}$. For tractability, we assume the noise precision $\rho$ follows a Gamma distribution
\begin{eqnarray}
p(\rho|a,b)=\mathcal{G}(\rho;a,b).\label{noisePrior2_MMV}
\end{eqnarray}
Therefore, the likelihood is given by
\begin{eqnarray}
p(\bm{X}|\bm{G},\rho)=\prod\limits_{l=1}^L\mathcal{CN}(\bm{x}_{\cdot l};\bm{A}\bm{g}_{\cdot l},\rho^{-1}\textbf{I}_M),\label{likelihood_MMV}
\end{eqnarray}
where $\bm{g}_{\cdot l}$ is the $l$'th column of $\bm{G}$. 

\subsubsection{Signal model}
Similarly to the SMV case, the MMV CAL-HAM is built with a three layer hierarchical model. We assume that the unknown signal $\bm{G}$ has group sparsity, i.e., each row of $\bm{G}$ shares a common parameter $\lambda_i$ that controls the sparsity. Thus, each row of $\bm{G}$ are modeled using a zero-mean multivariate Gaussian distribution, i.e.,    
\begin{eqnarray}
p(\bm{G}|\bm{\lambda})=\prod\limits_{i=1}^N p(\bm{g}_{i\cdot},\lambda_i)= \prod\limits_{i=1}^N \mathcal{CN}(\bm{g}_{i\cdot};\bm{0},\lambda_i\textbf{I}_L),\label{sigPrior1_MMV}
\end{eqnarray}
In order to assign adaptive Laplace priors to signals, the variables in $\bm{\lambda}$ are assumed to follow independent Gamma distributions with independent parameter $\gamma_i$, i.e.,
\begin{eqnarray}
p(\bm{\lambda}|\bm{\gamma})=\prod\limits_{i=1}^N p(\lambda_i|\gamma_i)= \prod\limits_{i=1}^N \mathcal{G}\bigg(\lambda_i;\frac{1}{2}+L,\frac{\gamma_i}{4}\bigg).\label{sigPrior2_MMV}
\end{eqnarray}
The variables in $\bm{\gamma}$ are assumed to follow independent Gamma distributions considering the conjugate prior rule, i.e., 
\begin{eqnarray}
p(\bm{\gamma}|c,d)=\prod\limits_{i=1}^N p(\gamma_i|c,d)=\prod\limits_{i=1}^N \Gamma(\gamma_i;c,d).\label{sigPrior3_MMV}
\end{eqnarray}

\begin{figure}[!t]
	\centering
	\includegraphics[width=4.5in]{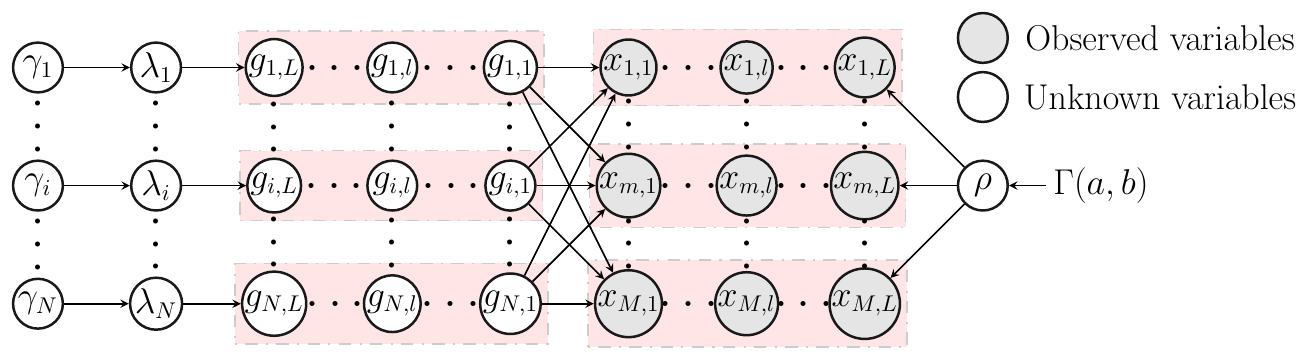}
	\caption{Directed acyclic graph representation for the hierarchical model with adaptive complex Laplace priors for the MMV case. The red rectangle indicate that the variables in the same row share a common parameter. }
	\label{fig:CAL-HBM-MMV}
\end{figure}

Similarly, the marginal distribution $p(\bm{G}|\bm{\gamma})$ can be calculated by considering the first two layers in this hierarchy. Thus, 
\begin{align}
&p(\bm{G}|\bm{\gamma})= \int p(\bm{G}|\bm{\lambda})p(\bm{\lambda}|\bm{\gamma}){\rm d}{\bm{\gamma}}\nonumber\\
&=  \int \prod\limits_{i=1}^N \mathcal{CN}(\bm{g}_{i\cdot };\bm{0},\lambda_i\bm{I}_L)\mathcal{G}\bigg(\lambda_i;\frac{1}{2}+L,\frac{\gamma_i}{4}\bigg){\rm d}{\bm{\gamma}}\nonumber\\
&=\frac{|\bm{\Lambda}|^L}{\bigg(4^L\sqrt{\pi}\Gamma\Big(\frac{1}{2}+L\Big)\bigg)^{N}}e^{-\sum\limits_{i=1}^N\sqrt{\gamma_i\left({\rm tr}(\bm{g}_{i\cdot }\bm{g}_{i\cdot }^{\rm H})\right)}}.\label{cLaplaceMMV}
\end{align}
It can be seen from (\ref{cLaplaceMMV}) that complex Laplace priors are assigned to signals in $\bm{G}$. Each row of $\bm{G}$ is controlled by a hyper-parameter $\gamma_i$. Consequently, the MMV CAL-HBM method is formulated by (\ref{likelihood_MMV}), (\ref{noisePrior2_MMV}), (\ref{sigPrior1_MMV}), (\ref{sigPrior2_MMV}) and (\ref{sigPrior3_MMV}), which corresponds to the MMV adaptive LASSO (\ref{ReWe_cLASSO_MMV}). To intuitively illustrate the structure of the proposed hierarchical Bayesian framework for the MMV case, Figure \ref{fig:CAL-HBM-MMV} is given as directed acyclic graph representation. The red rectangle indicate that variables in the same row share a common parameter.

\subsection{Relationship with state-of-the-art methods}
In this section, we discuss the relationship between the proposed hierarchical Bayesian model and state-of-the-art Bayesian models. The proposed CAL-HB model in viewpoint of maximum a posterior (MAP) estimation is given first. Then, the relationship between the proposed CAL-HB model and state-of-the-art hierarchical Bayesian models is discussed.

We use Type-I estimation approach to analysis the proposed CAL-HB model. In Type-I estimation approach, the MAP estimation of $\bm{G}$ is calculated as follows:
\begin{align}
\tilde{\bm{G}}&=\int\int p(\bm{X}|\bm{G})p(\bm{G}|\bm{\lambda})p(\bm{\lambda}|\bm{\gamma})p(\bm{\gamma})\text{d}\bm{\lambda}\text{d}\bm{\gamma}\nonumber\\
&=\int p(\bm{X}|\bm{G}) p(\bm{G}|\bm{\gamma})p(\bm{\gamma})\text{d}\bm{\gamma}. \label{TypeI0}
\end{align} 

According to (\ref{sigPrior3_MMV}) and (\ref{cLaplaceMMV}), the equation (\ref{TypeI0}) can be expressed as follows:
\begin{align}
\tilde{\bm{G}}^{\text{new}} =\arg\max\limits_{\bm{G}}\ln p(\bm{X}|\bm{G}) - \sum_{i=1}^{N}\int \gamma_i \|\bm{g}_{i\cdot}\|_2p\left(\gamma_i|\tilde{\bm{G}}^{\text{old}},c_i,d_i\right)\text{d}{\gamma_i}.  
\end{align}
The posterior of variable $\gamma_i$ follows a Gamma distribution, that is,
\begin{equation}
p\left(\gamma_i|\tilde{\bm{G}}^{\text{old}},c_i,d_i\right)=\mathcal{G}(\gamma_i|c_i+L,d_i+\|\bm{g}_{i\cdot}\|_2),
\end{equation}
resulting in 
\begin{equation}
\gamma_i=\frac{c_i+L}{d_i+\|\bm{g}_{i\cdot}\|_2}.
\end{equation}

By given a staring point $\tilde{\bm{G}}^0$, the global minima of the posterior $p(\bm{G}|\bm{X})$ can be find by solving
\begin{equation}
\tilde{\bm{G}}^{\text{new}}=\arg\min \rho\|\bm{X}-\bm{A}\bm{G}\|_2^2+\sum_{i=1}^{N}w_j^{\text{old}}\|\bm{g}_{i\cdot}\|_2,\label{typeI1}
\end{equation}
where $w_j^{\text{old}}=\frac{c_i+L}{d_i+\|\bm{g}_{i\cdot}\|_2}$.

It can be seen from (\ref{typeI1}) that the proposed hierarchical model corresponds to the adaptive LASSO framework, but the proposed model enjoys self-regularization and uncertainty nature.   

The hierarchical Bayesian framework proposed in \cite{Liu2012} and \cite{Tipping_2001}, i.e., so-called iRVM and  Stu-SBL, equals to build a one layer CAL-HB model and a two layer CAL-HB model, respectively. The hierarchical Bayesian framework using Laplace priors proposed in \cite{Babacan2010}, i.e., so-called Lap-SBL, equals to build a three layers CAL-HB model using a common parameter in the third layer. The difference between the proposed CAL-HB model and Lap-SBL model is that the proposed CAL-HB model assigns independent Gamma priors to the precision parameters of Gaussian distributions, resulting in adaptive Laplace priors. Besides, the proposed MMV CAL-HBM is a general extension of the proposed SMV CAL-HBM. 

%In order to intuitively show the cause of sparsity, we plot the contours of cost function of joint priors for two dimensional signal. Figure \ref{sparsityGragh}  shows the contours of different priors, i.e., iRVM denotes sparse priors proposed in \cite{Liu2012}, Stu-SBL denotes Student-t priors proposed in \cite{Tipping_2001}, Lap-SBL denotes Laplace priors proposed in \cite{Babacan2010}, CAL-HB denotes the proposed adaptive Laplace priors. It can be seen from Figure \ref{sparsityGragh} that the contour of the proposed adaptive Laplace priors are more sharply peaked at zero and more closely to the axis comparing with other contours, indicating that the proposed adaptive Laplace priors encourage sparsity more efficiently than state-of-the-art sparse priors.  
%\begin{figure}[htbp]
%	\centering
%	\includegraphics[width=2in]{./Fig/SparsityGraph.pdf}
%	\caption{Two dimensional contour plots of different priors}
%	\label{sparsityGragh}
%\end{figure} 

\section{Variational Bayesian inference}\label{sec:VBI}
Thus far, we have presented the proposed CAL-HB model. However, it is intractable to calculate the true posterior distribution. Therefore, the variational Bayesian inference is used to calculate all hidden parameters in CAL-HB model. To further reduce computational complexity, the space alternative method is utilized to avoid matrix inverse operation. For the convenience of comprehension, the derivation of the SMV case is given first. Then, the MMV case is given as an extension of the SMV case.

\subsection{Variational Bayesian inference}
In variational Bayesian inference, the true posterior distribution $p(\bm{\theta}|\bm{x})$ is approximated by a distribution $q(\bm{\theta})$, which has a factorized form as follows:
\begin{equation}
q(\bm{\theta})=q_g(\bm{g})q_{\lambda}(\bm{\lambda})q_{\gamma}(\bm{\gamma})q_{\rho}(\rho),
\end{equation}
where $\bm{\theta}=\{\bm{g},\bm{\lambda},\bm{\gamma},\rho\}$ is the set of all unknown parameters. 
The logarithmic evidence $\ln p(\bm{x})$ can be written as
\begin{eqnarray}
\ln p(\bm{x})= \int q(\bm{\theta})\ln\frac{ p(\bm{x},\bm{\theta})}{q(\bm{\theta})}{\rm d}\bm{\theta}+\int q(\bm{\theta})\ln\frac{q(\bm{\theta})}{ p(\bm{\theta}|\bm{x})}{\rm d}\bm{\theta}.\label{LBandKL}
\end{eqnarray}
The first and second terms on the right side of (\ref{LBandKL}) are the evidence lower bound $\mathcal{L}$ and the Kullback–Leibler (KL) divergence $\mathcal{K}$, respectively, i.e.,
\begin{equation}
\mathcal{L}=\int q(\bm{\theta})\ln\frac{ p(\bm{x},\bm{\theta})}{q(\bm{\theta})}{\rm d}\bm{\theta},\;
\mathcal{K}=\int q(\bm{\theta})\ln\frac{q(\bm{\theta})}{     p(\bm{\theta}|\bm{x})}{\rm d}\bm{\theta}.
\end{equation}

The parameters in the approximate distribution $q(\bm{\theta})$ are calculated by minimizing the KL divergence $\mathcal{K}$. Because the evidence $p(\bm{x})$ is a constant and $\mathcal{K}\geqslant 0$, minimizing the KL divergence is equivalent to maximizing the lower bound $\mathcal{L}$, which results in \cite{Tzikas2008}
\begin{eqnarray}
&\ln q(\bm{\theta}_k)=  {\rm E}_{q(\bm{\theta}\backslash\bm{\theta}_k)}\big(\ln p(\bm{x},\bm{\theta})\big)+ {\rm{cons}},\label{VBI}\\
&p(\bm{x},\bm{\theta})=  p(\bm{x}|\bm{g},\rho)p(\bm{g}|\bm{\lambda})p(\bm{\lambda}|\bm{\gamma})p(\bm{\gamma})p(\rho), \label{JointDist}
\end{eqnarray}
where $\rm{cons}$ denotes a constant, $\bm{\theta}_k$ denotes a subset of $\bm{\theta}$, and $\bm{\theta}\backslash\bm{\theta}_k$ denotes the subset of $\bm{\theta}$ with $\bm{\theta}_k$ pruned.

\subsection{Bayesian inference for the SMV case}
Substituting the likelihood (\ref{obserPrio}), the Gamma prior of noise precision (\ref{noisePrior2}), the priors of the unknown signal (\ref{sigPrior1}), (\ref{sigPriorReWe2}) and (\ref{sigPriorReWe3}) into (\ref{JointDist}), the joint distribution can be written as follows:
\begin{eqnarray}
&\ln p(\bm{x},\bm{\theta})=M\ln\rho-\rho\left\|\bm{x}-\bm{A}\bm{g}\right\|^2+\sum\limits_{i=1}^N\ln\lambda_i^{-1}-\bm{g}^{\rm H}{\bm{\Lambda}}^{-1}\bm{g}\nonumber\\
&+\frac{3}{2}\sum\limits_{i=1}^N\ln\gamma_i+\frac{1}{2}\sum\limits_{i=1}^N\ln\lambda_i-\frac{1}{4}\sum\limits_{i=1}^N\gamma_i\lambda_i+\sum\limits_{i=1}^N(c-1)\ln\gamma_i-\nonumber\\
&\sum\limits_{i=1}^N d\gamma_i+(a-1)\ln\rho-b\rho+c_x,
\end{eqnarray}
where $c_x$ denotes the normalization factor constant of the joint distribution. In the following derivation, $c_g$, $c_{\lambda_i}$, $c_{\gamma_i}$ and $c_{\rho}$ denote the normalization factor constant of each distribution, respectively. Given the form of joint distribution, all of the parameters can be updated using (\ref{VBI}) and (\ref{JointDist}), as shown in the following.

\paragraph{Update of $\bm{g}$}
According to (\ref{VBI}), the variational approximation of $\ln p(\bm{g}|\bm{x})$ is 
\begin{align}
\ln q(\bm{g})=& {\rm E}_{q(\bm{\theta}\backslash\bm{g})}\big(\ln p(\bm{x},\bm{\theta})\big)\nonumber\\
&=  -\bm{g}^{\rm H}\Big({\rm{E}}_{q(\rho)}(\rho)\bm{A}^{\rm H}\bm{A}+{\rm{E}}_{q(\bm{\lambda})}(\bm{\Lambda}^{-1})\Big)\bm{g}+\nonumber\\
&{\rm{E}}_{q(\rho)}(\rho)\bm{g}^{\rm H}\bm{A}^{\rm H}\bm{x}+
{\rm{E}}_{q(\rho)}(\rho)\bm{x}^{\rm H}\bm{A}\bm{g}+c_{g},
\end{align}
which indicates that $q(\bm{g})$ can be described using a multivariate complex Gaussian distribution with the mean and variance given as
\begin{align}
\bm{\mu}_{g}&= {\rm E}_{q(\rho)}(\rho)\bm{\Sigma}\bm{A}^{\rm H}\bm{x},\nonumber\\
\bm{\Sigma}_{g}&= \Big({\rm E}_{q(\rho)}(\rho)\bm{A}^{\rm H}\bm{A}+{\rm E}_{q(\bm{\lambda})}\big(\bm{\Lambda}^{-1}\big)\Big)^{-1}.
\end{align}
Therefore, we have the following update rule for $\bm{g}$:
\begin{eqnarray}
{\rm E}_{q(\bm{g})}(\bm{g})={\rm E}_{q(\rho)}(\rho)\bm{\Sigma}\bm{A}^{\rm H}\bm{x}.\label{update_gJoint}
\end{eqnarray}

\paragraph{Update of $\bm{\lambda}$, $\bm{\gamma}$ and $\rho$ }
See Appendix \ref{Appendix_B} 

The unknown parameters can be updated iteratively using (\ref{update_gJoint}), (\ref{update_lambda}), (\ref{update_lambdaInv}), (\ref{update_gamma}) and (\ref{update_rho}). However, the matrix inverse operation is required in (\ref{update_gJoint}), which entails a heavy computational load that may be prohibitive for some applications. To reduce the computational complexity, the space alternative strategy is used in the variational Bayesian inference, as detailed next.

\subsubsection{Space alternative variational estimation}
If we assume that each element in $\bm{g}$ is independent and the approximate posterior $q(\bm{g})$ can be factorized, then we have $q(\bm{g})=\prod\limits_{i=1}^N q(g_i)$. Thus, the approximate posterior of $g_i$ can be derived from (\ref{VBI}). That is,
\begin{align}
&\ln q(g_i)= {\rm E}_{q(\bm{\theta}\backslash g_i)}\left(\ln p(\bm{x},\bm{\theta})\right)\nonumber\\
&=  g_i^{\rm H}\bigg({\rm E}_{q(\rho)}(\rho)\bm{A}_i^{\rm H}\bm{A}_i+{\rm E}_{q(\lambda_i)}\big(\lambda_i^{-1}\big)\bigg)g_i-{\rm E}_{q(\rho)}(\rho)g_i^{\rm H}\bm{A}_i^{\rm H}\times\nonumber\\
&\bigg(\bm{x}-\bm{A}_{ \bar{i}}{\rm E}_{q(\bm{g}_{ \bar{i}})}\big(\bm{g}_{ \bar{i}}\big)\bigg)-
{\rm E}_{q(\rho)}(\rho)\bigg(\bm{x}-\bm{A}_{ \bar{i}}{\rm E}_{q(\bm{g}_{ \bar{i}})}\big(\bm{g}_{ \bar{i}}\big)\bigg)^{\rm H}\times\nonumber\\
&\bm{A}_i g_i+{c_{g_i}},\label{post_g_factor}
\end{align}
where $\bm{A}_i$ denotes the $i$'th column of the dictionary $\bm{A}$, $g_i$ denotes the $i$'th element in the vector $\bm{g}$, $\bm{A}_{ \bar{i}}$ denotes a sub-dictionary of $\bm{A}$ with the $i$'th column pruned, and $\bm{g}_{ \bar{i}}$ denotes the vector with the $i$'th element pruned. Note that an equality formulation is used in (\ref{post_g_factor}), i.e.,
\begin{equation}
\bm{A}\bm{x}=\bm{A}_i g_i+\bm{A}_{ \bar{i}}\bm{g}_{ \bar{i}}.
\end{equation}

Considering the quadratic form of (\ref{post_g_factor}), a complex Gaussian distribution can be used to represent the approximate posterior $q(g_i)$ with following parameters: 
\begin{align}
\sigma_i^2&=  \frac{1}{{\rm E}_{q(\rho)}(\rho)\bm{A}_i^{\rm H}\bm{A}_i+{\rm E}_{q(\lambda_i)}\big(\lambda_i^{-1}\big)},\label{update_gVar}\\
\mu_i&=  \sigma_i^2{\rm E}_{q(\rho)}(\rho)\bm{A}_i^{\rm H}\bigg(\bm{x}-\bm{A}_{ \bar{i}}{\rm E}_{q(\bm{g}_{ \bar{i}})}\big(\bm{g}_{ \bar{i}}\big)\bigg).\label{update_gMean}
\end{align}
Thus, the update rule of $g_i$ is
\begin{equation}
{\rm E}_{q(g_i)}=\sigma_i^2{\rm E}_{q(\rho)}(\rho)\bm{A}_i^{\rm H}\bigg(\bm{x}-\bm{A}_{ \bar{i}}{\rm E}_{q(\bm{g}_{ \bar{i}})}\big(\bm{g}_{ \bar{i}}\big)\bigg).
\end{equation}
In this way, all the hidden parameters in the proposed CAL-HB model are calculated using the space alternative variational estimation algorithm. Specifically, the unknown parameters can be updated using (\ref{update_gMean}), (\ref{update_lambda}), (\ref{update_lambdaInv}), (\ref{update_gamma}) and (\ref{update_rho}), iteratively. For simplicity, we refer the proposed algorithm as CAL-SAVE.  

\subsubsection{Update the hyper-parameters $a$ and $b$}\label{update_abcd}
To achieve a low computational complexity performance, the hyper-parameters $a$ and $b$ are fixed in the algorithm above. However, if we choose to update these hyper-parameters, then the recovery accuracy performance will be improved when the number of the measurement is sufficiently high, which is a trade-off between the performance of the computational complexity and the recovery accuracy. Because there is no posterior for the hyper-parameters $a$ and $b$, these parameters are updated by maximizing the joint distribution $\ln p(\bm{x},\bm{\theta})$.

The derivative of $\ln p(\bm{x},\bm{\theta})$ with respect to $a$ and $b$ is given by
\begin{align}
\frac{\partial \ln p(\bm{x},\bm{\theta})}{\partial a}&= \ln b-\Psi(a)+{\rm E}_{q(\rho)}(\ln\rho),\label{deri_a}\\
\frac{\partial \ln p(\bm{x},\bm{\theta})}{\partial b}&= \frac{a}{b}-{\rm E}_{q(\rho)}(\rho),\label{deri_b}
\end{align}
where $\Psi(\cdot)$ denotes the digamma function.
The update rule of $b$ can be derived from (\ref{deri_b}), i.e., 
\begin{equation}
b=\frac{a}{{\rm E}_{q(\rho)}(\rho)}.
\end{equation}
However, there is no closed-form solution for (\ref{deri_a}), and the hyper-parameter must be updated numerically. 

\subsection{Bayesian inference for the MMV case}
Similarly to the SMV case, the joint distribution $p(\bm{X},\bm{G},\bm{\lambda},\bm{\gamma},\rho)$ can be represented using the likelihood (\ref{likelihood_MMV}), the Gamma prior of noise precision (\ref{noisePrior2_MMV}), the priors of unknown signal (\ref{sigPrior1_MMV}), (\ref{sigPrior2_MMV}) and (\ref{sigPrior3_MMV}). As a result, we obtain 
\begin{align}
&\ln p(\bm{X},\bm{\theta})=\ln \big(p(\bm{X}|\bm{G},\rho)p(\bm{G}|\bm{\lambda})p(\bm{\lambda}|\bm{\gamma})p(\bm{\gamma}|c,d)p(\rho|a,b)\big)\nonumber\\
&=LM\ln{\rho}-\rho\left\|\bm{X}-\bm{A}\bm{G}\right\|_f^2+L\sum\limits_{i=1}^N\ln{\lambda_i^{-1}}-
\sum\limits_{i=1}^N\lambda_i^{-1}{\rm{tr}}\left(\bm{g}_{i\cdot}^{\rm H}\bm{g}_{i\cdot}\right)\nonumber\\
&+\big(\frac{1}{2}+L\big)\sum\limits_{i=1}^N\ln{\gamma_i}+\big(L-\frac{1}{2}\big)\sum\limits_{i=1}^N\ln{\lambda_i}-\frac{1}{4}\sum\limits_{i=1}^N\gamma_i\lambda_i+\nonumber\\
&(c-1)\sum\limits_{i=1}^N\ln\gamma_i-d\sum\limits_{i=1}^N\gamma_i+(a-1)\ln\rho-b\rho+c_X,\label{evi_MMV}
\end{align}
where $c_X$ is a constant of the normalized factor. In the following, $c_G$, $c_{\lambda}$, $c_{\gamma}$ and $c_{\rho}$ represent the constants of the normalized factor for the MMV case. 

\paragraph{Update of $\bm{g}_{i\cdot}$}
Because we assume that the unknown signal $\bm{G}$ has group sparsity, each row can be processed independently using the SAVE algorithm. According to (\ref{VBI}) and (\ref{evi_MMV}), the approximate posterior of $\bm{g}_{i\cdot}$ is
\begin{align*}
&\ln q(\bm{g}_{i\cdot})= \ln {\rm E}_{q(\bm{\theta}\backslash \bm{g}_{i\cdot})}\big(p(\bm{X},\bm{\theta})\big)\nonumber\\
&= -{\rm tr}\bigg(\bm{g}_{i\cdot}^{\rm H}\big({\rm E}_{q(\rho)}(\rho)\bm{A}_i^{\rm H}\bm{A}_i+{\rm E}_{q(\lambda_i)}\big(\lambda_i^{-1}\big)\big)\bm{g}_{i\cdot}+\nonumber\\
&{\rm E}_{q(\rho)}(\rho)\bm{g}_{i\cdot}^{\rm H}\bm{A}_i^{\rm H}\big(\bm{X}-\bm{A}_{ \bar{i}}{\rm E}_{q({\bm{G}}_{ \bar{i\cdot}})}({\bm{G}}_{ \bar{i\cdot}})\big)+\nonumber\\
&{\rm E}_{q(\rho)}(\rho)\big(\bm{X}-\bm{A}_{ \bar{i}}{\rm E}_{q({\bm{G}}_{ \bar{i\cdot}})}({\bm{G}}_{ \bar{i\cdot}})\big)^{\rm H}\bm{A}_i \bm{g}_{i\cdot}\bigg)+c_G,
\end{align*}
where $\bm{g}_{i\cdot}$ denotes the $i$'th row of the signal $\bm{G}$, and ${\bm{G}}_{ \bar{i\cdot}}$ denotes the signal matrix with the $i$'th row removed.
The multivariate quadratic form of $\ln q(\bm{g}_{i\cdot})$ indicates that the approximate distribution $q(\bm{g}_{i\cdot})$ can be described using a multivariate complex Gaussian distribution with parameters given as
\begin{align}
\sigma_{i}^2&= \frac{1}{{\rm E}_{q(\rho)}(\rho)\bm{A}_i^{\rm H}\bm{A}_i+{\rm E}_{q(\lambda_i)}\big(\lambda_i^{-1}\big)},\label{update_sigma_MMV}\\
\bm{\mu}_{i\cdot}&= \sigma_{i}^2{\rm E}_{q(\rho)}(\rho)\bm{A}_i^{\rm H}\big(\bm{X}-\bm{A}_{ \bar{i}}{\rm E}_{q({\bm{G}}_{ \bar{i\cdot}})}({\bm{G}}_{ \bar{i\cdot}})\big).\label{mu_MMV}
\end{align}
Therefore, $\bm{g}_{i\cdot}$ is updated using
\begin{equation}
{\rm E}_{q(\bm{g}_{i\cdot})}(\bm{g}_{i\cdot})=\sigma_{i}^2{\rm E}_{q(\rho)}(\rho)\bm{A}_i^{\rm H}\big(\bm{X}-\bm{A}_{ \bar{i}}{\rm E}_{q({\bm{G}}_{ \bar{i\cdot}})}({\bm{G}}_{ \bar{i\cdot}})\big).\label{update_mu_MMV}
\end{equation}

\paragraph{Update of $\bm{\lambda}$, $\bm{\gamma}$ and $\rho$ }
See Appendix \ref{Appendix_MMV} 

\begin{algorithm}
\SetAlgoLined
\KwIn{The MMV data $\bm{X}$, the dictionary $\bm{A}$, and the parameters $a$, $b$, $c$ and $d$.}
\KwOut{The recovered signal matrix $\bm{G}$ and the variance matrix $\bm{\Sigma}$.}
Initialize the precision of noise $\rho=\frac{a}{b}$\;
Initialize the hyper-parameter $\gamma_i=\frac{c}{d}$, $\gamma_i^{-1}=\frac{d}{c}$, and $\lambda_i=\frac{6}{\gamma_i}$, for $i=1,\cdots,N$\;
Initialize the mean of the variable $\bm{G}=\bm{0}$ and the variance of the variable $\sigma_i^2=\frac{1}{\rho\bm{A}_i^{\rm H}\bm{A}_i+\lambda_i^{-1}}$, for $i=1,\cdots,N$\;
Initialize the temporary variable $\bm{X}_{\rm tmp}=AG$\;  
\While{convergence criterion not met}{
	\For{$i \gets 1$ \bf{to} $N$} {
		Update $\bm{g}_{i\cdot}^{\rm old} \gets \bm{g}_{i\cdot}$\;
		Update $\bm{X}_{\rm tmp} \gets \bm{X}_{\rm tmp}-\bm{A}_i\bm{g}_{i\cdot}^{\rm old}$\;
		Update $\sigma^2_i \gets \frac{1}{\rho\bm{A}_i^{\rm H}\bm{A}_i+\gamma_i^{-1}}$ using (\ref{update_sigma_MMV})\;
		Update $\bm{g}_{i\cdot} \gets \sigma_{i}^2\rho\bm{A}_i^{\rm H}\big(\bm{X}-\bm{X}_{\rm tmp}\big)$ using (\ref{update_mu_MMV})\;
		Update $\bm{X}_{\rm tmp} \gets \bm{X}_{\rm tmp}+\bm{A}_i\bm{g}_{i\cdot}$\;
		Update $\lambda_i \gets 2\bigg(\frac{\sqrt{{\rm{tr}}(\bm{g}_{i\cdot}^{\rm H}\bm{g}_{i\cdot})+\sigma^2_i}}{\sqrt{\gamma_i}}+\frac{1}{\gamma_i}\bigg)$ using (\ref{update_lambda_MMV})\;
		Update $\lambda_i^{-1} \gets \frac{1}{\lambda_i-\frac{2}{\gamma_i}}$ using (\ref{update_lambdaInv_MMV})\;
		Update $\gamma_i \gets \frac{4(L+c)+2}{\lambda_i+4d}$ using (\ref{update_gamma_MMV})\;
	}
	Update $\rho \gets \frac{LM+a}{\left\|\bm{X}-\bm{A}{\rm E}_{q(\bm{G})}(\bm{G})\right\|^2+
		\sum\limits_{i=1}^N\sigma_i^2\bm{A}_i^{\rm H}\bm{A}_i+b}$ based on (\ref{update_rho_MMV})\; 
	%			\eIf{update $a$ and $b$}{
		%				Update $a$ and $b$ based on Section \ref{update_abcd}\;
		%			}{
		%				Fix $a$ and $b$ to the initial values\;
		%			}
}
\caption{An overview of the proposed algorithms}
\label{alg:cSAVE_SBL}
\end{algorithm}

All of the hidden parameters are updated using (\ref{update_mu_MMV}), (\ref{update_lambda_MMV}), (\ref{update_lambdaInv_MMV}), (\ref{update_gamma_MMV}) and (\ref{update_rho_MMV}), sequentially. The proposed algorithm for the MMV case is summarized in Algorithm \ref{alg:cSAVE_SBL}\footnote{To avoid duplicate calculation, a temporary variable $\bm{X}_{\rm tmp}$ is used in the algorithm. Note that the SMV CAL-SAVE is a special case of the MMV CAL-SAVE, where the number of measurement vectors is $1$.}.

\subsection{Computational complexity analysis}
The computational complexity is calculated by counting the number of multiplications/divisions and additions. As can be seen from Algorithm \ref{alg:cSAVE_SBL}, the computational complexity of the proposed algorithm is caused mainly by updating the signal $\bm{g}_{i\cdot}$ in each `for' loop. The computational complexity of the proposed method is thus $\mathcal{O}(MNL)$ for each iteration. The complexity complexity of other SBL methods using the Woodbury identity is $\mathcal{O}(MN^2)$. It should be remarked that in most cases, $L$ is smaller than $N$. Also, for some applications such as DOA estimation, the singular value decomposition (SVD) can be applied to reduce the amount of data, i.e., $L$ is a small number comparing with $N$. As a result, the proposed method can be applied to reduce the computational complexity when $L<N$ or the SVD can be used for data dimensional reduction beforehand. Note that the proposed method achieves excellent recovery performance even in the single measurement vector case.    

\section{Experimental results}\label{sec:exp}

In this section, we first test the recovery accuracy of different algorithms with complex Gaussian random dictionaries. Then, acoustic DOA estimation application is considered. We compare the performance of the proposed method with other state-of-the-art methods that are widely used in sparse signal recovery. All the methods in the comparison are summarized as follows:
%	All computations are performed on a 3.20 GHz Intel Core i7 8700 series 8 core machine with 16 GB of RAM. The MATLAB version is R2019a.
\begin{itemize}
\item `CAL-SAVE' refers to the proposed method based on the hierarchical Bayesian model using complex adaptive Laplace priors	\footnote{The MATLAB code for the proposed algorithm is available online: https://tinyurl.com/ub5jroa}.
\item `CL-SAVE' refers to the SAVE based method based on the hierarchical model with a common $\eta$ parameter, as proposed in \cite{Thomas2018}.
\item `MSBL' is an SBL method using student-t priors for complex-value signal recovery, as proposed in \cite{Gerstoft2016}.
\item `l1-l2' refers to a new $\ell_1-\ell_2$ minimization algorithm  proposed in \cite{Zihao2022}.
\item `TISTA' refers to a trainable iterative soft thresholding algorithm proposed in \cite{Daisuke2019}.
\item `FLap-Real' is a Laplace signal model-based fast SBL method that uses the basis addition and deletion strategy proposed in \cite{Babacan2010}. 
\item `LASSO' refers to the LASSO method based on the complex-value signal model \cite{Tibshirani1996}.
\item `MFOCUSS' refers to the focal under-determined system solver proposed in \cite{Cotter2005a}. Following the setup in the paper, the parameter $p$ is set to $0.8$.
\item `CSMUSIC' is a subspace pursuit method for sparse signal recovery, as proposed in \cite{Dai2009}.
\item `KF-RTSS-SBL' is an SBL method using the Kalman filter and Rauch Tung Striebel smoother to accelerate the algorithm, as proposed in \cite{8410591}.
\item `EM-VB' is a fast SBL method using a basis addition and deletion strategy proposed in \cite{9110823},  and the Bayesian model is built based on the hierarchical Bayesian framework in \cite{Ji2008}.
\item `NOMP' is  the newtonized orthogonal matching pursuit method, as proposed in \cite{7491265} and further improved in \cite{9181410}.

\end{itemize}

\subsection{Experimental setup}

To quantify the sparse recovery performance of different algorithms, the normalized mean square error (NMSE) criteria is used, defined as
\begin{eqnarray}
e_{\rm NMSE}=10\log_{10}\left(\frac{1}{LN_{MC}}\sum_{t=1}^{N_{MC}}\frac{\left\|\tilde{\bm{G}}^{t}-\bm{G}^{t}\right\|_f^2}{\|\bm{G}^{t}\|_f^2}\right),
\end{eqnarray}
where $t$ and $N_{MC}$ are index and total numbers of Monte-Carlo experiments, respectively. Moreover, $K$ is the number of non-zero rows in $\bm{G}$ and $L$ is the number of measurement vectors. $\tilde{\bm{G}}^t$ and ${\bm{G}}^t$ are the estimation and the true value in the $t$'th Monte-Carlo experiment, respectively. 

We test the performance of all methods in different scenarios, more specifically for different SNR, number of non-zero elements $K$ and length of measurement $M$, respectively. For each case, three types of signal are used for testing, i.e., complex Gaussian signals, complex Laplace signals and complex spike signals. The real and image parts of these signals are generated independently following the same distribution with a common variance value. The dictionaries are built by sampling from a normal distribution, i.e., $\Re(a_{i,j}) \sim \mathcal{N}(0,1)$ and $\Im(a_{i,j}) \sim \mathcal{N}(0,1)$, and are normalized for each row. The index of non-zero elements is selected randomly. The parameters $a$, $b$, $c$ and $d$ are set to a small value, e.g., $10^{-6}$. For all of the cases, the number of Monte-Carlo experiments is set to $1000$. We test the performance for the SMV and MMV cases in order.

\subsection{Performance analysis for the SMV case}

The experiment setup for the SMV case is summarized as follows:
\begin{itemize}
\item In the first experiment, the recovery accuracy performance of different methods is tested versus the length of measurement $M$. More specifically, $M$ ranges from $50$ to $150$ with an interval of $10$. The length of signal $N$ is fixed to $200$ while the number of non-zero elements $K$ is set to $20$. The SNR is set to $10$ dB.
\item In the second experiment, the recovery accuracy performance is tested versus different numbers of non-zero elements $K$. The length of signal $N$ is fixed to $200$, while $K$ changes from $5$ to $30$ with an interval of $5$. The length of measurement $M$ is fixed to $100$ and the SNR is set to $10$ dB.
\item In the third experiment, we test the recovery accuracy performance of different methods versus different SNRs. The SNR ranges from $0$ dB to $30$ dB with an interval of $5$ dB. The length of measurement $M$ is set to $100$. The length of signal $N$ is set to $200$, and the number of non-zero elements is set to $K=20$.
\end{itemize}	

\begin{figure*}[!t]
	\centering
	\subfloat[]{\includegraphics[width=0.3\textwidth]{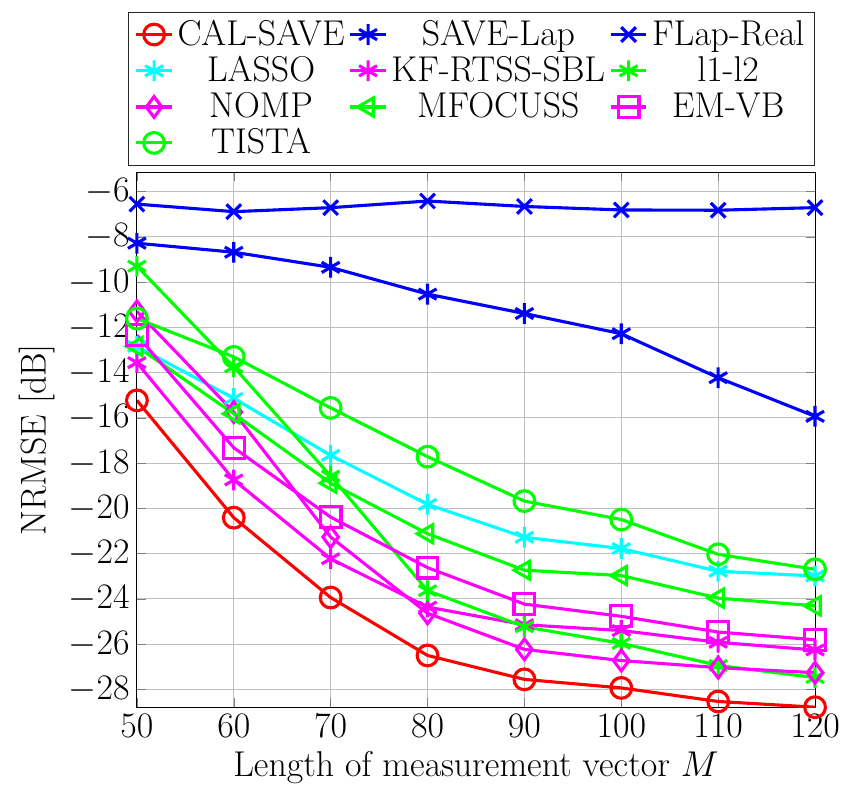}
		\label{fig_MMSE_vsMeasurements_GaussianSig}}
	\hfil
	\subfloat[]{\includegraphics[width=0.3\textwidth]{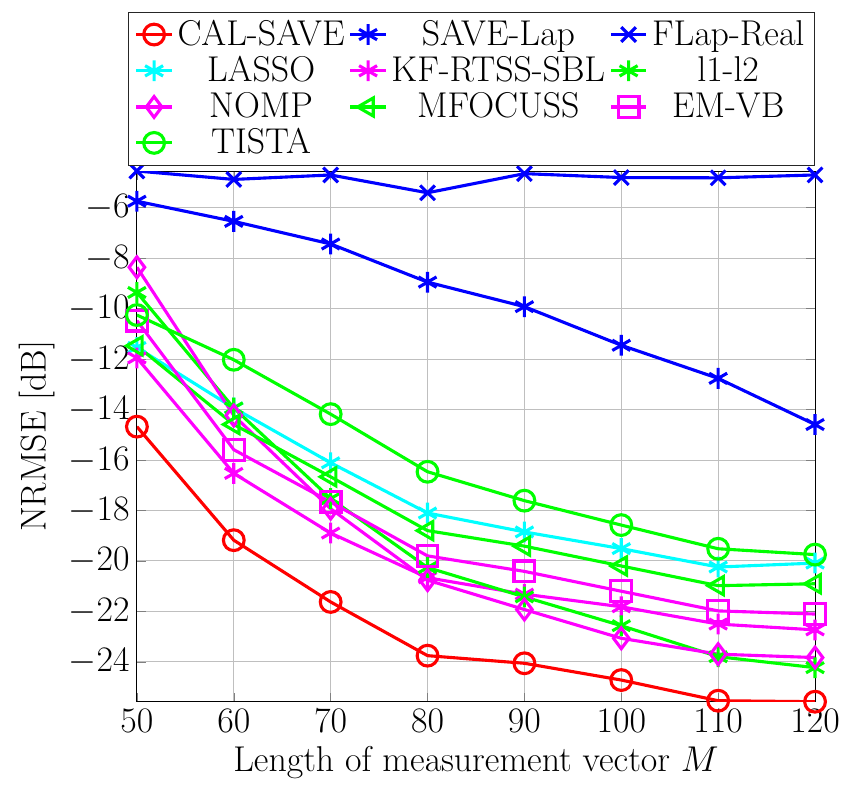}
		\label{fig_MMSE_vsMeasurements_LapSig}}
	\hfil
	\subfloat[]{\includegraphics[width=0.3\textwidth]{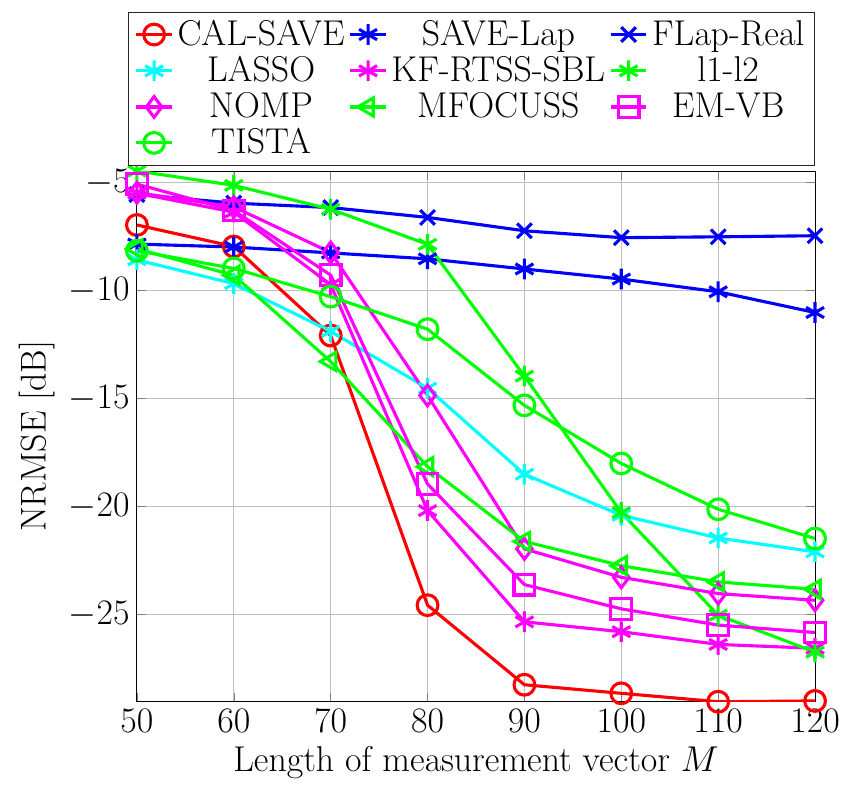}
		\label{fig_MMSE_vsMeasurements_SpikeSig}}
	\vfil
	\subfloat[]{\includegraphics[width=0.3\textwidth]{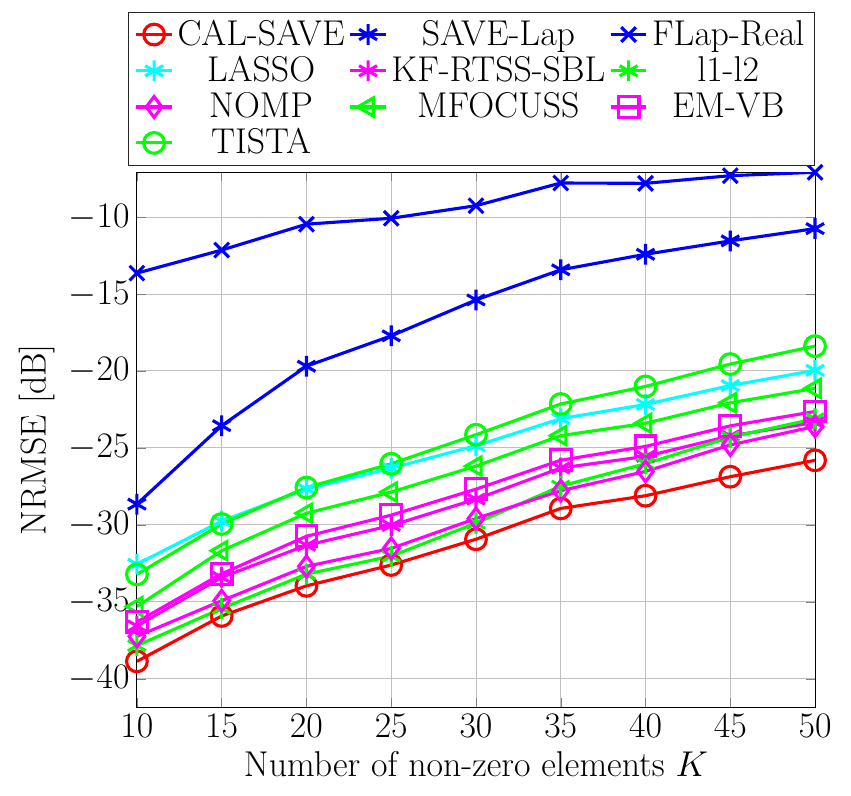}
		\label{fig_MMSE_vsNumSources_GaussianSig}}
	\hfil
	\subfloat[]{\includegraphics[width=0.3\textwidth]{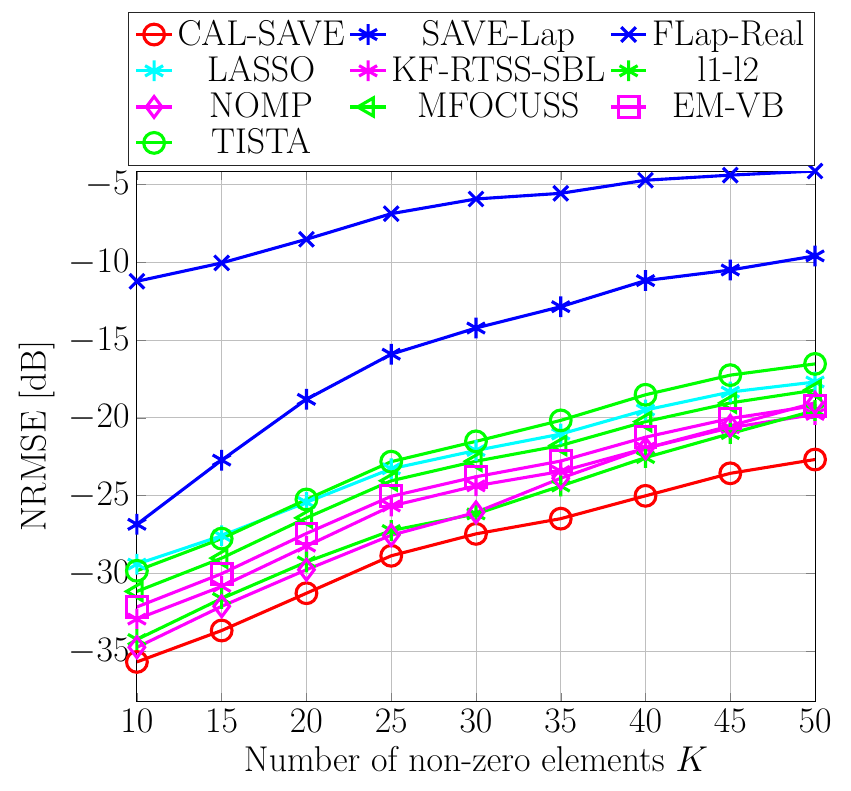}
		\label{fig_MMSE_vsNumSources_LapSig}}
	\hfil
	\subfloat[]{\includegraphics[width=0.3\textwidth]{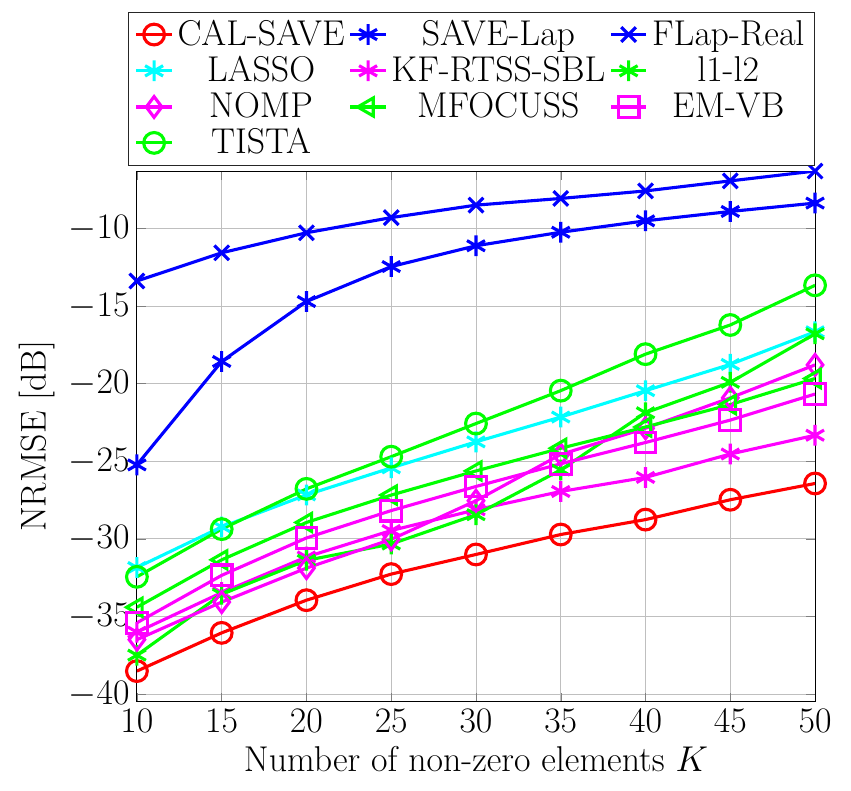}
		\label{fig_MMSE_vsNumSources_SpikeSig}}
	\vfil
	\subfloat[]{\includegraphics[width=0.3\textwidth]{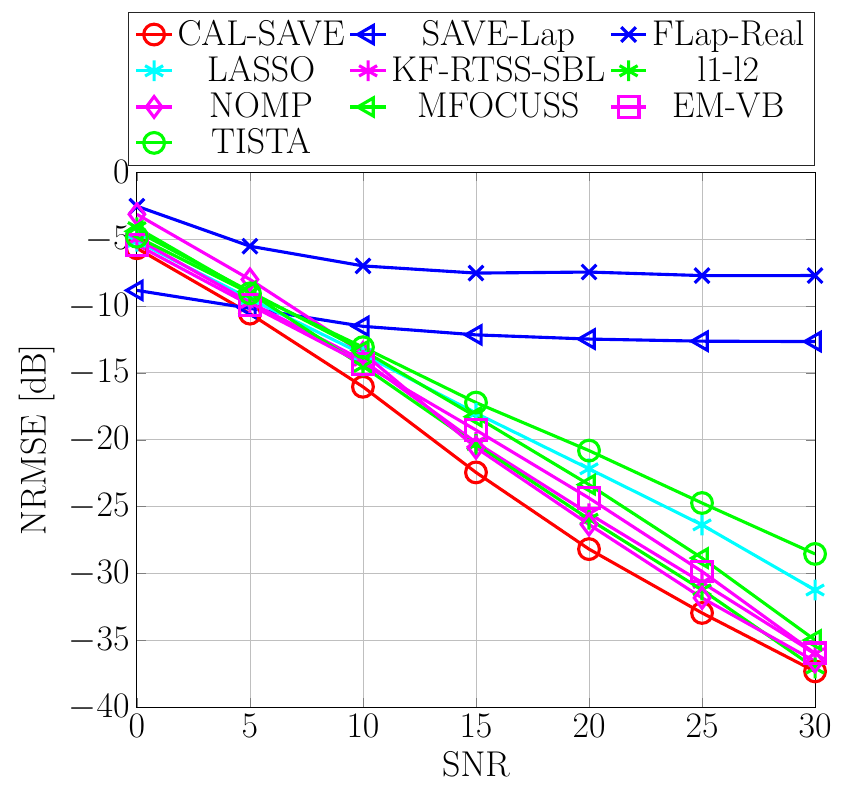}
		\label{fig_MMSE_vsSNRs_GaussianSig}}
	\hfil
	\subfloat[]{\includegraphics[width=0.3\textwidth]{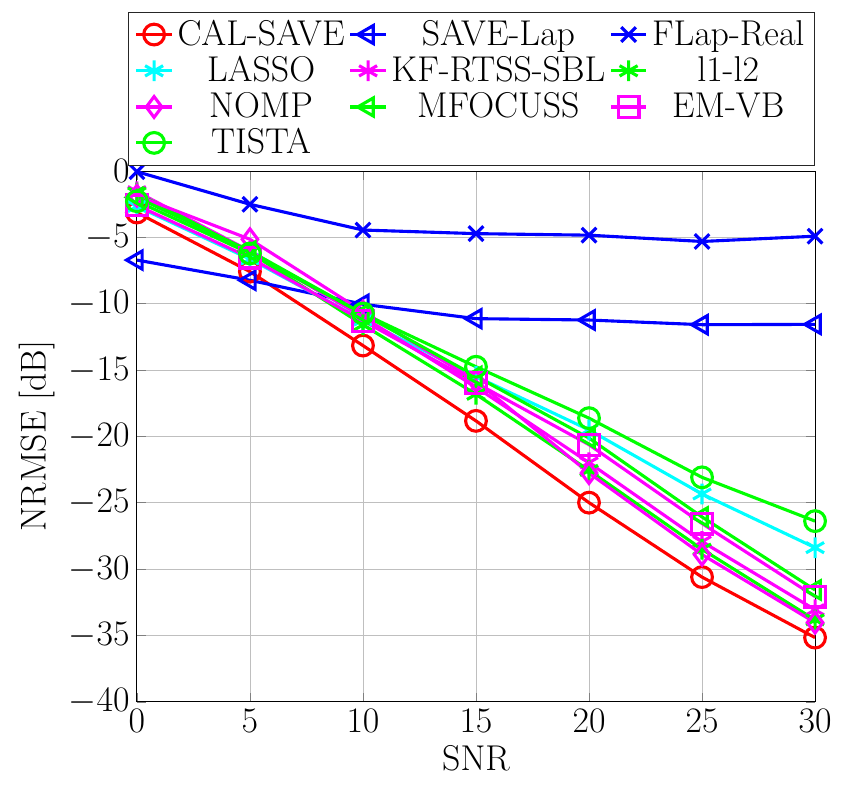}
		\label{fig_MMSE_vsSNRs_LapSig}}
	\hfil
	\subfloat[]{\includegraphics[width=0.3\textwidth]{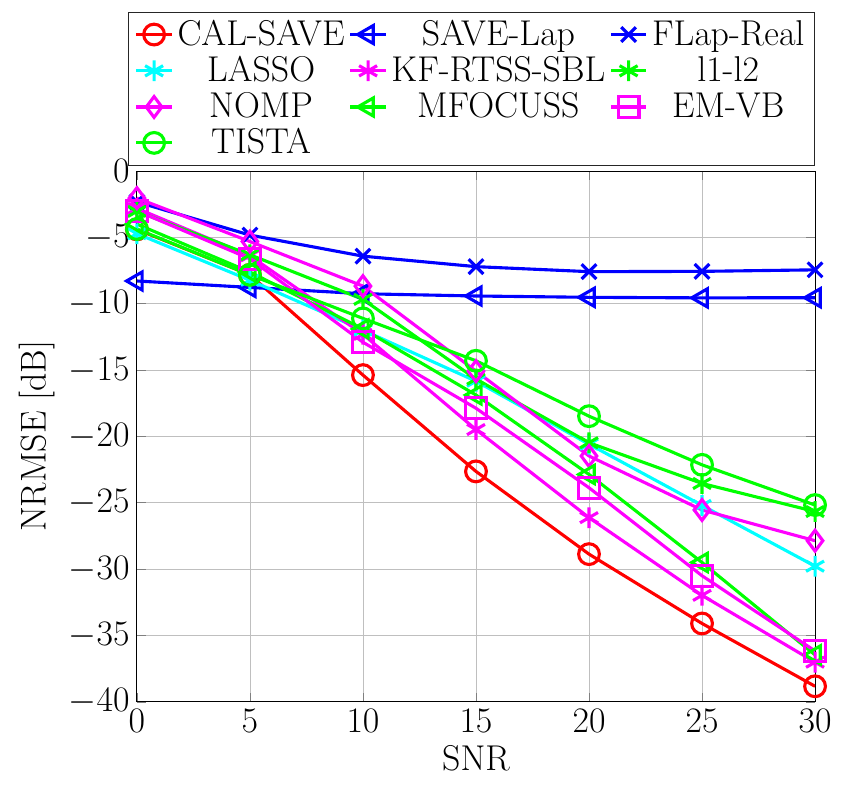}
		\label{fig_MMSE_vsSNRs_SpikeSig}}
	\caption{Simulation results for the SMV case using complex Gaussian random dictionaries. Specifically, (a), (b) and (c) are NMSEs versus different length of measurement $M$ for Gaussian signals, Laplace signals and spike signals, respectively; (d), (e) and (f) are NMSEs versus different numbers of non-zero elements $K$ for Gaussian signals, Laplace signals and spike signals, respectively; (g), (h) and (i) are NMSEs versus different SNRs for Gaussian signal, Laplace signals and spike signals, respectively.}
	\label{fig_RandMatrix_SMV}
\end{figure*}

Figure \ref{fig_MMSE_vsMeasurements_GaussianSig}, Figure \ref{fig_MMSE_vsMeasurements_LapSig} and Figure \ref{fig_MMSE_vsMeasurements_SpikeSig} show the recovery accuracy of each algorithm versus $M$ for complex Gaussian signals, complex Laplace signals and complex spike signals, respectively. It can be seen that the proposed `CAL-SAVE' achieves the best performance for all cases. The reason for this result is that the statistical model of the proposed approach matches the sparsity nature of the signals. In contrast to the `CAL-SAVE' method, both the `CL-SAVE' and `FLap-Real' methods are assigned with a common Gamma prior in the last layer of the hierarchical Bayesian framework, resulting in bad recovery performance in this case. The difference between these two methods is that `CL-SAVE' is derived from the variational Bayesian inference, whereas `FLap-Real' is based on evidence maximization (type-II maximum likelihood) \cite{Bishop2006}. The `KF-RTSS-SBL' and `EM-VB' are all fast SBL methods based on the typical hierarchical Bayesian framework using student-t priors. Specifically, the `KF-RTSS-SBL' method reduces computational complexity using KF and RTSS, and the `EM-VB' method  reduces computational complexity using a basis addition and deletion strategy to avoid matrix inversion.  As can be seen from Figure \ref{fig_MMSE_vsMeasurements_GaussianSig}, Figure \ref{fig_MMSE_vsMeasurements_LapSig} and Figure \ref{fig_MMSE_vsMeasurements_SpikeSig}, the proposed CAL-SAVE method achieves higher recovery accuracy performance than state-of-the-art methods. 

The recovery accuracy results versus different sparsity levels are illustrated in Figure \ref{fig_MMSE_vsNumSources_GaussianSig}, Figure \ref{fig_MMSE_vsNumSources_LapSig} and Figure \ref{fig_MMSE_vsNumSources_SpikeSig}, respectively. As can be seen from these figures, the performance of all methods is degraded with an increasing of number of non-zero elements $K$. Note that a suitable regularization factor is required for the l1-l2 algorithm and it achieves a better recovery performance than the other methods in the range from $10$ to $20$. However, the proposed `CAL-SAVE' outperforms the other methods when the number of non-zero elements is larger than $20$, without prior knowledge of the number of non-zero elements.

Figure \ref{fig_MMSE_vsSNRs_GaussianSig}, Figure \ref{fig_MMSE_vsSNRs_LapSig} and Figure \ref{fig_MMSE_vsSNRs_SpikeSig} show the recovery performance under different SNRs for the complex Gaussian signal, complex Laplace signal and complex spike signal, respectively. The recovery accuracy of all methods increases as SNR increases. In comparison with other state-of-the-art methods, the proposed `CAL-SAVE' method has the best recovery accuracy performance in the range from $5$ to $30$ dB.

\subsection{Performance for the MMV case}

\begin{figure*}
\centering
\begin{subfigure}[b]{0.3\textwidth}
	\centering
	\includegraphics[width=1\textwidth]{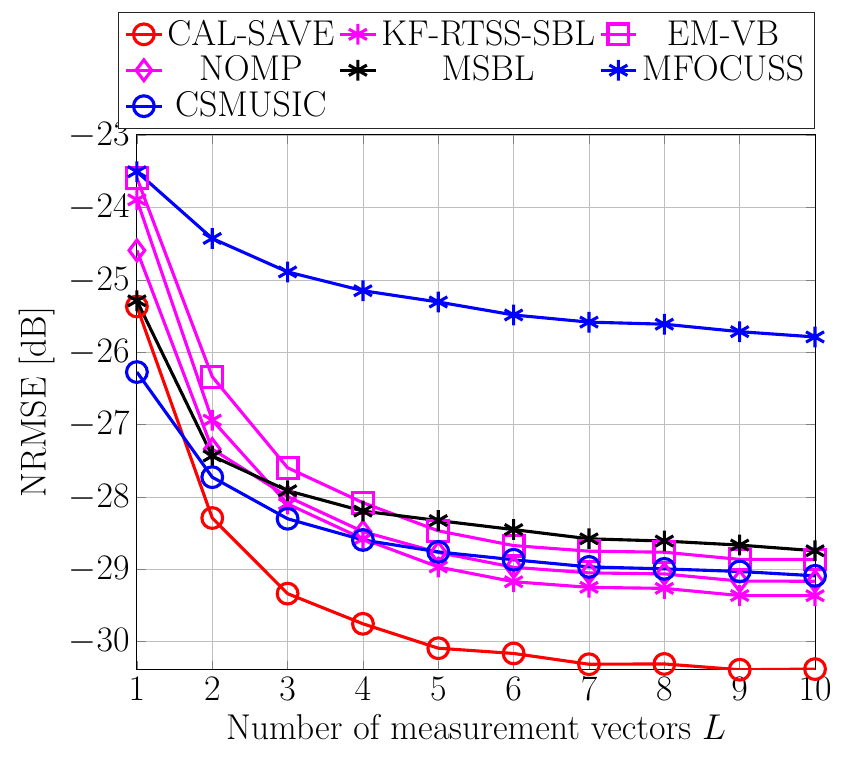}
	\caption{}
	\label{fig_MMV_vsNumSnap_GaussianSig}
\end{subfigure}
\hfil
\begin{subfigure}[b]{0.3\textwidth}
	\centering
	\includegraphics[width=1\textwidth]{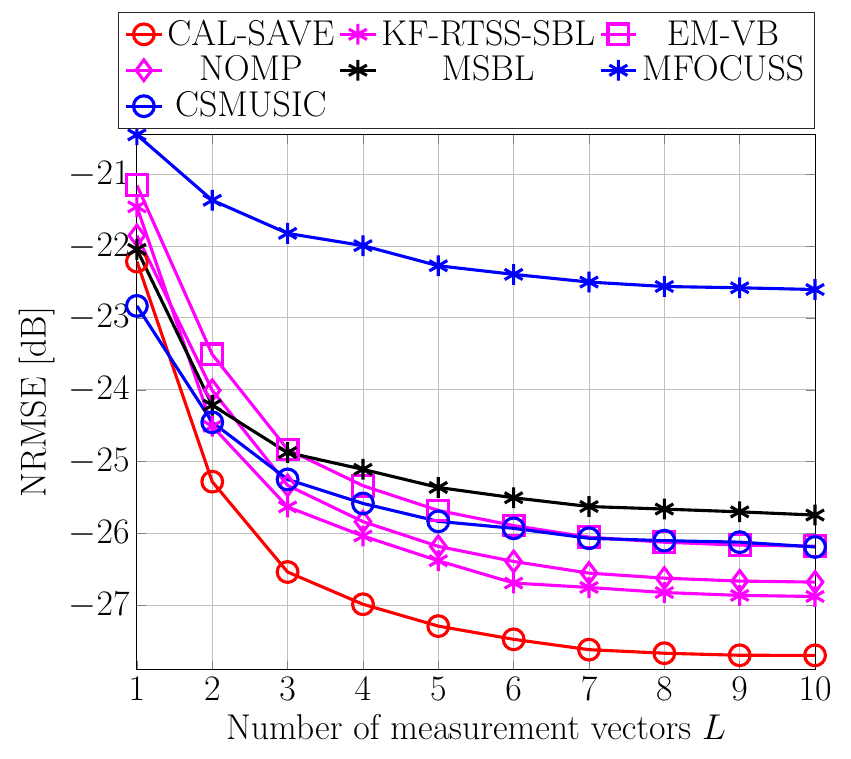}
	\caption{}
	\label{fig_MMV_vsNumSnap_LapSig}
\end{subfigure}
\hfil
\begin{subfigure}[b]{0.3\textwidth}
	\centering
	\includegraphics[width=1\textwidth]{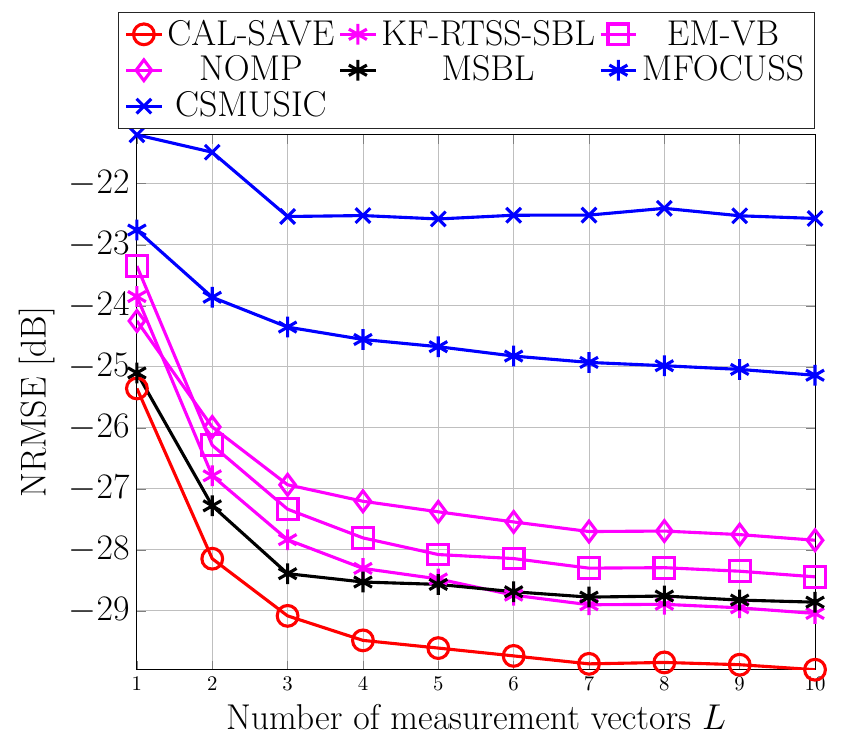}
	\caption{}
	\label{fig_MMV_vsNumSnap_SpikeSig}
\end{subfigure}
\vfil
\begin{subfigure}[b]{0.3\textwidth}
	\centering
	\includegraphics[width=1\textwidth]{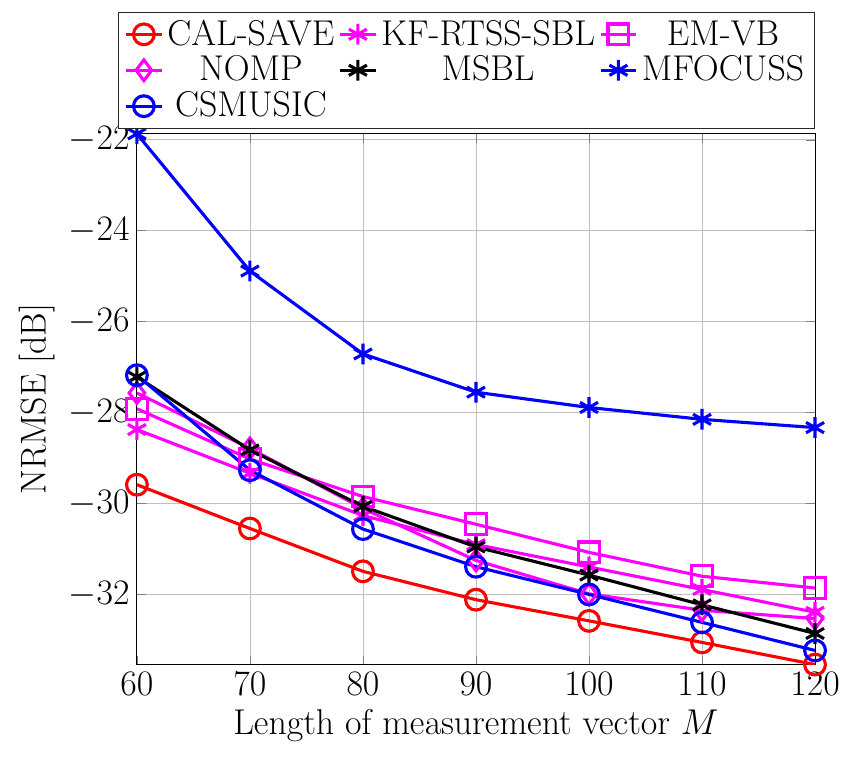}
	\caption{}
	\label{fig_MMV_vsMeasurements_GaussianSig}
\end{subfigure}
\hfil
\begin{subfigure}[b]{0.3\textwidth}
	\centering
	\includegraphics[width=1\textwidth]{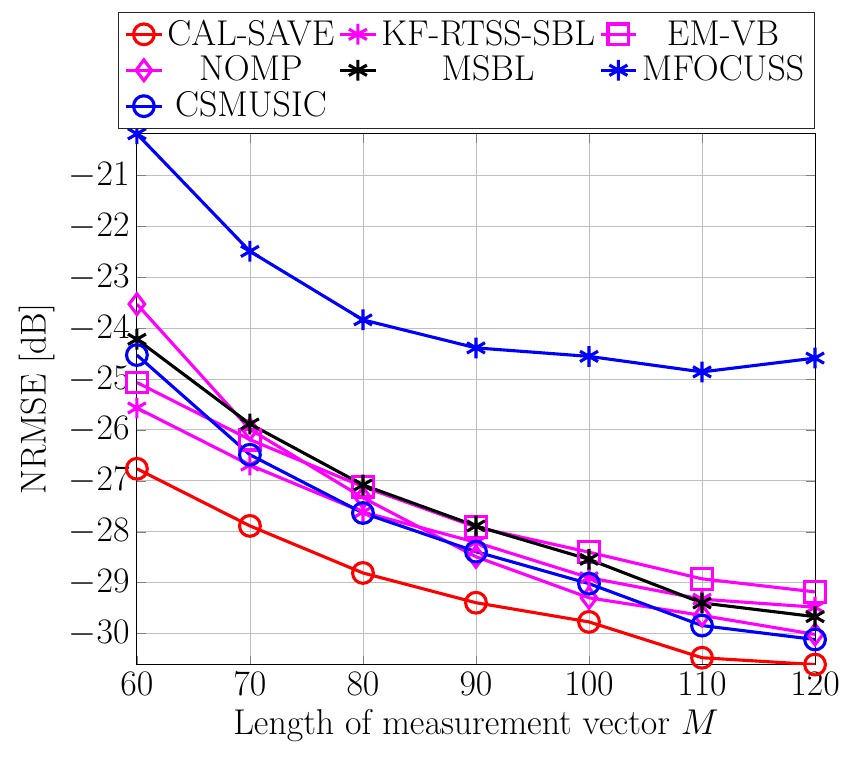}
	\caption{}
	\label{fig_MMV_vsMeasurements_LapSig}
\end{subfigure}
\hfil
\begin{subfigure}[b]{0.3\textwidth}
	\centering
	\includegraphics[width=1\textwidth]{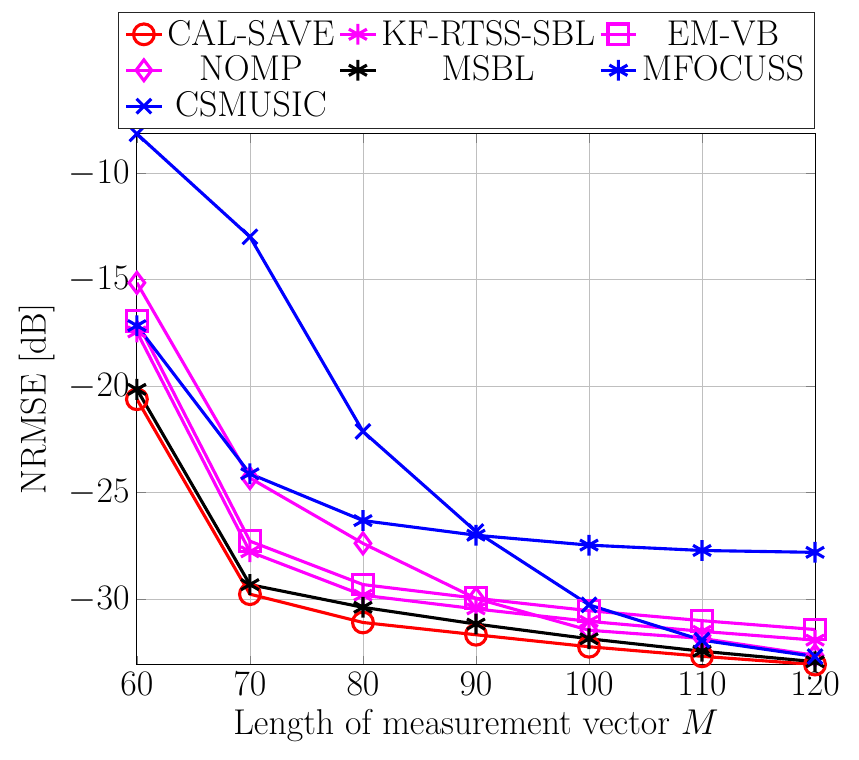}
	\caption{}
	\label{fig_MMV_vsMeasurements_SpikeSig}
\end{subfigure}
\vfil
\begin{subfigure}[b]{0.3\textwidth}
	\centering
	\includegraphics[width=1\textwidth]{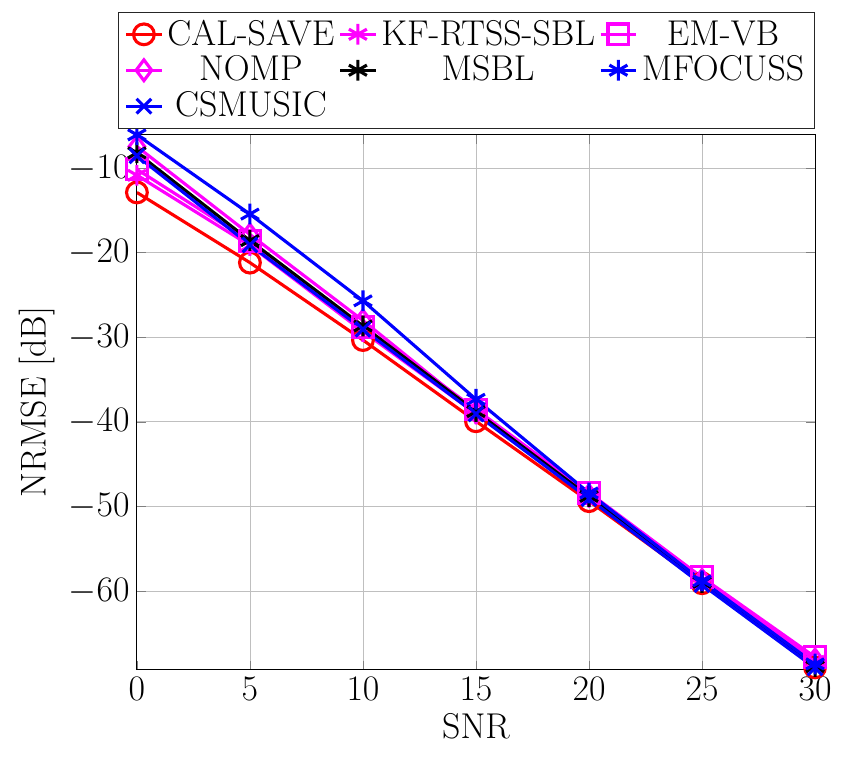}
	\caption{}
	\label{fig_MMV_vsSNRs_GaussianSig}
\end{subfigure}
\hfil
\begin{subfigure}[b]{0.3\textwidth}
	\centering
	\includegraphics[width=1\textwidth]{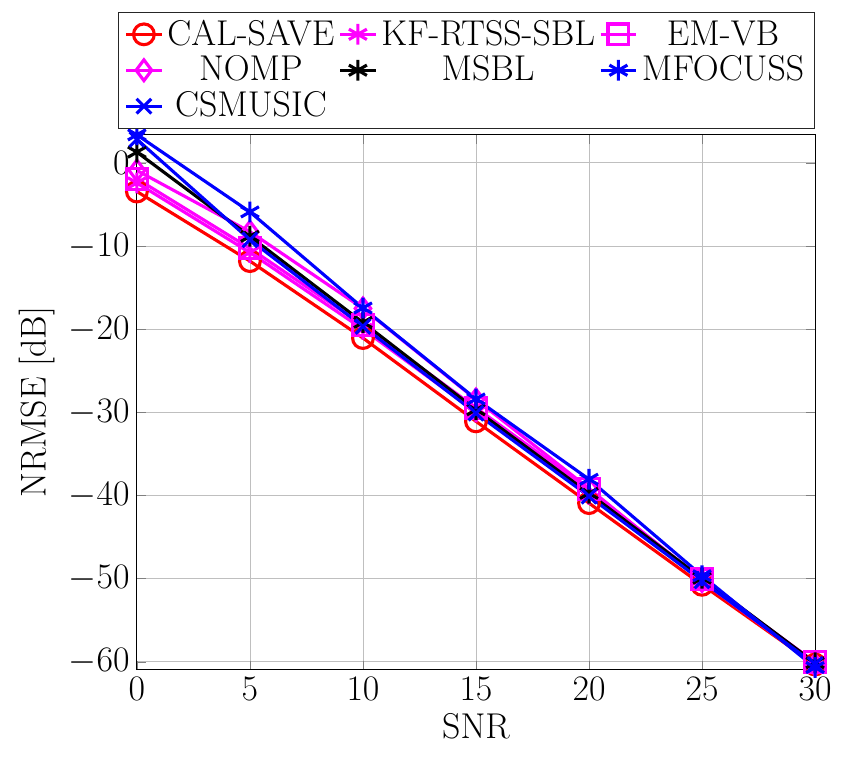}
	\caption{}
	\label{fig_MMV_vsSNRs_LapSig}
\end{subfigure}
\hfil
\begin{subfigure}[b]{0.3\textwidth}
	\centering
	\includegraphics[width=1\textwidth]{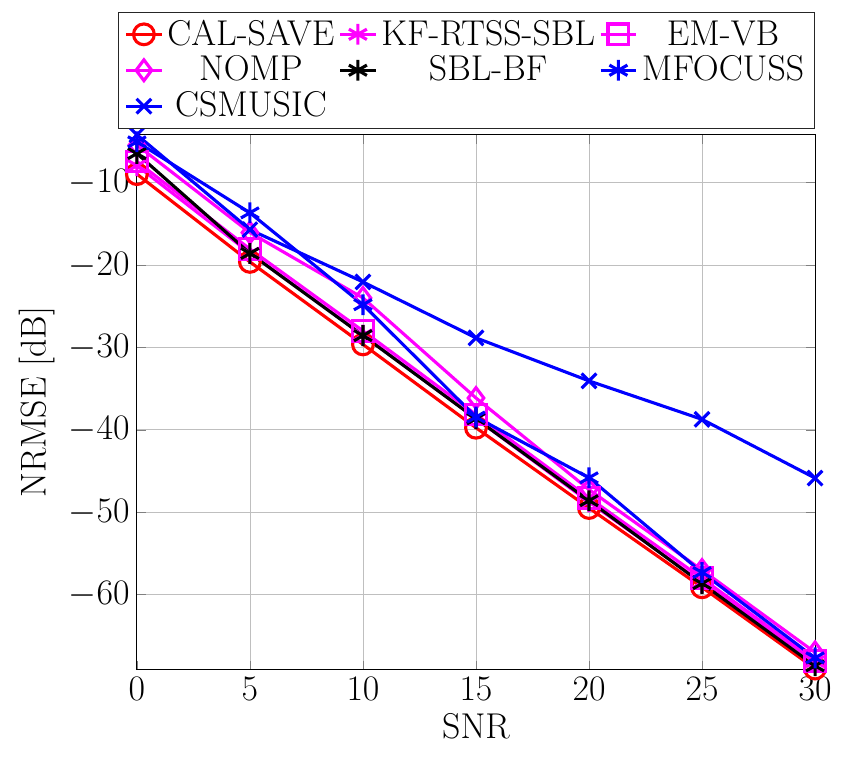}
	\caption{}
	\label{fig_MMV_vsSNRs_SpikeSig}
\end{subfigure}
\vfil
\begin{subfigure}[b]{0.3\textwidth}
	\centering
	\includegraphics[width=1\textwidth]{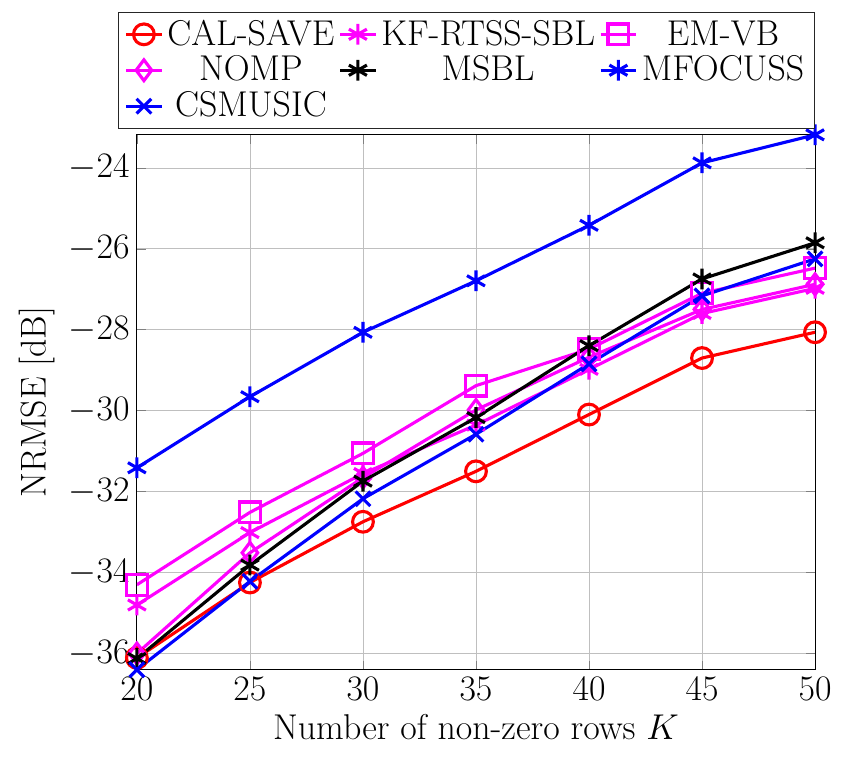}
	\caption{}
	\label{fig_MMV_vsSpar_GaussianSig}
\end{subfigure}
\hfil
\begin{subfigure}[b]{0.3\textwidth}
	\centering
	\includegraphics[width=1\textwidth]{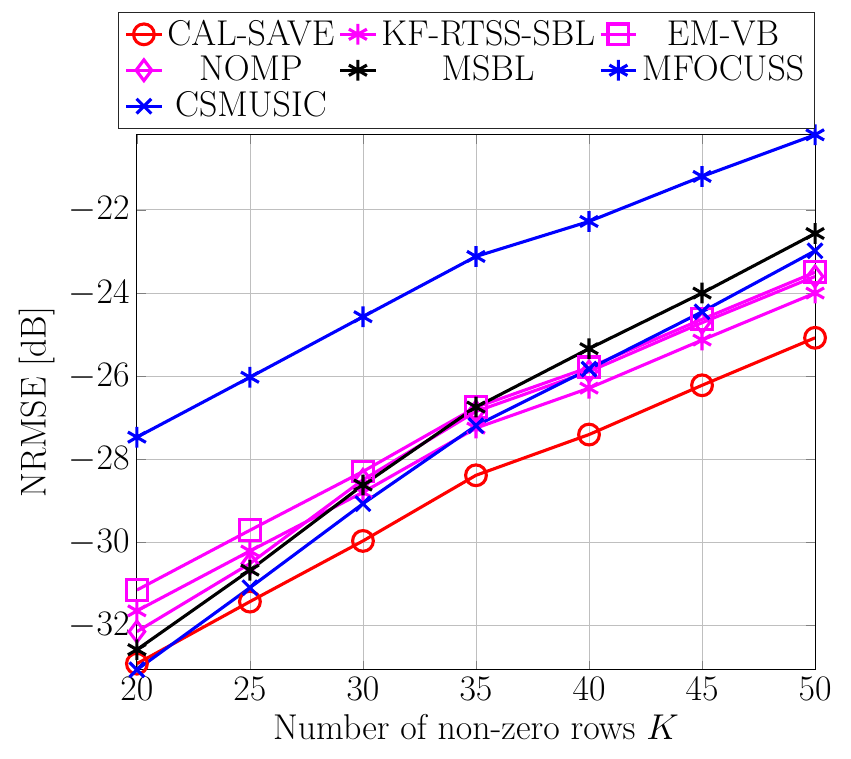}
	\caption{}
	\label{fig_MMV_vsSpar_LapSig}
\end{subfigure}
\hfil
\begin{subfigure}[b]{0.3\textwidth}
	\centering
	\includegraphics[width=1\textwidth]{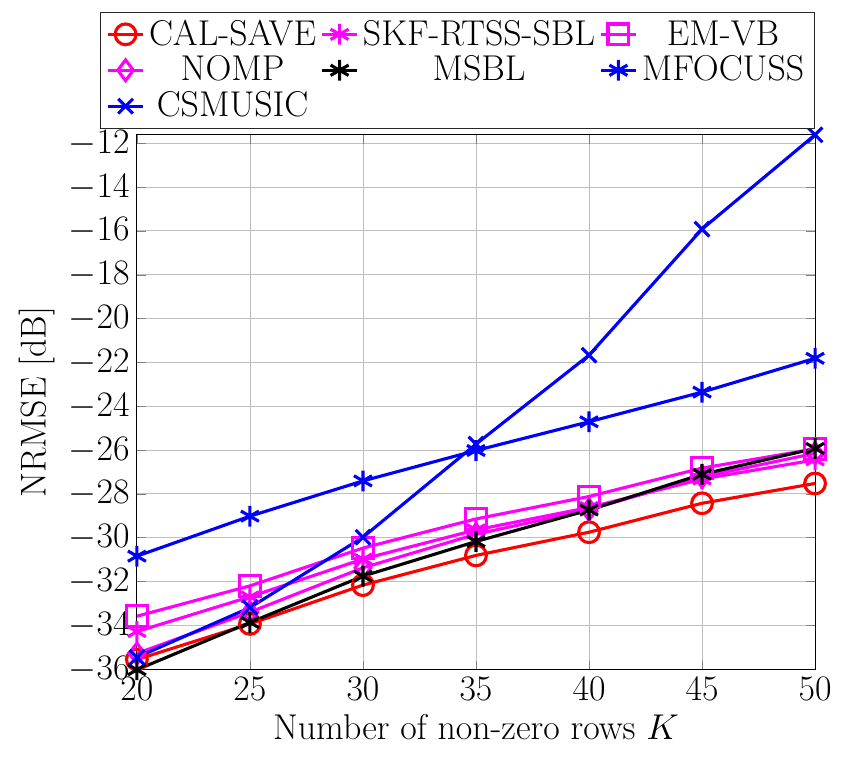}
	\caption{}
	\label{fig_MMV_vsSpar_SpikeSig}
\end{subfigure}
\caption{Simulation results for the MMV case using complex Gaussian random dictionaries. Specifically, (a), (b) and (c) are NMSEs versus different numbers of measurement vectors $L$ for Gaussian signals, Laplace signals and spike signals, respectively; (d), (e) and (f) are NMSEs versus different length of measurement $M$ for Gaussian signals, Laplace signals and spike signals, respectively; (g), (h) and (i) are NMSEs versus SNRs for Gaussian signals, Laplace signals and spike signals, respectively; (j), (k) and (l) are NMSEs versus different numbers of non-zero rows $K$ for Gaussian signals, Laplace signals and spike signals, respectively.}
\label{fig_RandMatrix_MMV}
\end{figure*}

For the MMV case, the recovery accuracy performance of the proposed method is verified as follow.
In the first experiment, the recovery performance is tested versus different numbers of measurement vectors $L$ in the range from $1$ to $10$ with an interval of $1$. The length of measurement $M$ is fixed to $100$. The length of signal $N$ is fixed to $200$, while the number of non-zero rows $K$ is set to $40$. The SNR is set to $10$ dB. In the next three simulations, we test the recovery performance versus different $M$, SNR and $K$, successively. The setup is identical to the SMV case except that the number of measurement vectors $L$ is set to $5$. 

In the first experiment, the recovery performance of different methods is tested for different number of measurements $L$, as shown in Figure \ref{fig_MMV_vsNumSnap_GaussianSig}, Figure \ref{fig_MMV_vsNumSnap_LapSig} and Figure \ref{fig_MMV_vsNumSnap_SpikeSig}. It can be seen from these figures that using a larger number of measurement vectors leads to a better recovery performance, as expected. For the `KF-RTSS-SBL' method and `EM-VB' method, they are all based on the SBL framework using student-t priors. As a result, the recovery accuracy performance of `KF-RTSS-SBL' method and `EM-VB' method is close to the typical MSBL algorithm. For the `CSMUSIC' methods, we assume that the sparsity is known. As a result, the performance of the `CSMUSIC' method is better than that of the others when $L=1$. Compared to the `CSMUSIC' methods, the proposed `CAL-SAVE' method does not require prior knowledge of the sparsity level but does achieve a better recovery performance than the state-of-the-art methods when $L$ is greater than $1$.

Figure \ref{fig_MMV_vsMeasurements_GaussianSig}, Figure \ref{fig_MMV_vsMeasurements_LapSig} and Figure \ref{fig_MMV_vsMeasurements_SpikeSig} show the recovery performance versus length of measurement $M$ for complex Gaussian signals, complex Laplace signals and complex spike signals, respectively. Similarly to the SMV case, longer length of measurement $M$ leads to better recovery performance. As can be seen from these figures, the proposed method achieves a better performance than the `MFOCUSS', `MSBL', `KF-RTSS-SBL', `EM-VB' and `CSMUSIC' methods in most of scenarios while requires no prior knowledge of the sparsity level. 

The recovery performance versus different SNRs for complex Gaussian signals, complex Laplace signals and complex spike signals is tested in the third experiment, as shown in Figure \ref{fig_MMV_vsSNRs_GaussianSig}, Figure \ref{fig_MMV_vsSNRs_LapSig} and Figure \ref{fig_MMV_vsSNRs_SpikeSig}, respectively. The proposed `CAL-SAVE' method exploits the sparsity nature of signals with an adaptive framework. As a result, it outperform the others in the range from $0$ dB to $20$ dB. 

Figure \ref{fig_MMV_vsSpar_GaussianSig}, Figure \ref{fig_MMV_vsSpar_LapSig} and Figure \ref{fig_MMV_vsSpar_SpikeSig} show the recovery accuracy performance of different methods versus different numbers of non-zero rows $K$ for complex Gaussian signals, complex Laplace signals and complex spike signals, respectively. Both the `MSBL' and `CSMUSIC' methods perform better under high and known sparsity levels, i.e., small number of non-zero rows $K$. However, similar to the SMV case, the proposed method achieves a better performance when the number of non-zero rows $K$ is greater than $25$, even without prior knowledge of the sparsity level.  

\subsection{Acoustic DOA estimation}
\begin{figure}[htbp]
\centering
\includegraphics[width=2in]{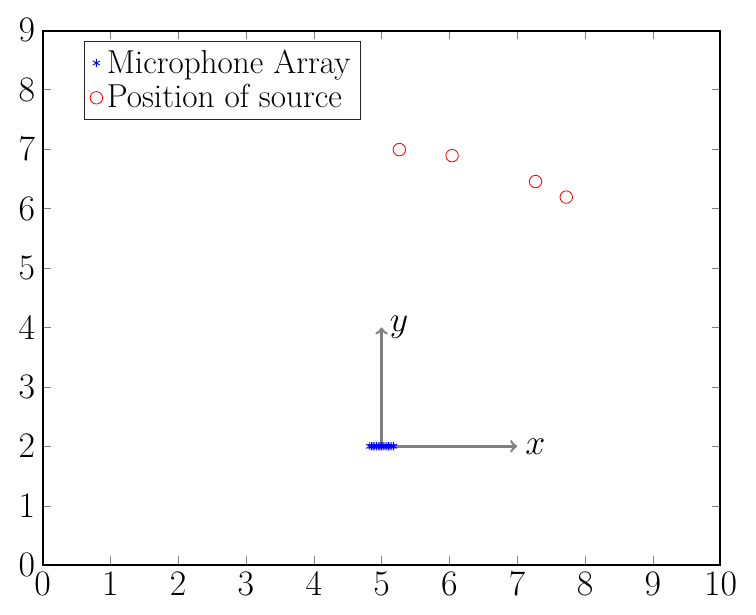}
\caption{Room setup for acoustic DOA estimation.}
\label{fig_room_setup}
\end{figure}

In this subsection, the proposed method is applied to acoustic DOA estimation to improve the resolution performance by encouraging the sparsity of sources in spatial domain. To verify the high resolution performance of the proposed algorithm for acoustic DOA estimation, we carry out an experiment using `RIR-generator' software\cite{rir_generator}. Figure \ref{fig_room_setup} gives an illustration of room setup, where blue stars denote microphones and red circles denote acoustic sources.  
In this experiment, the room dimension, the reverberation time RT60, the reverberation impulse response (RIR) length, the reflection order and sound speed are set to $10\times9\times8$m, $0.25$s, $2048$, $3$ and $343$m/s, respectively. Moreover, an uniform linear microphone array is used to receive array signals. The number of microphones is set to $15$, and the interval between two adjacent microphones is set to $0.025$m. The 
center point of the microphone array is set to $(5,2,1)$m. We define the horizontal plane in front of the microphone array as the target plane, the center point of microphone array as the reference point, the positive direction of $x$ axis as $-90^\circ$, the positive direction of $y$ axis as $0^\circ$ and the negative direction of $x$ axis is $90^\circ$, respectively. The sampling frequency is set to $16$kHz. Four pure sinusoidal acoustic sources are used in this experiment. The frequency values of four sources are all set to $1$kHz and the initial phase values of four sources are set randomly. The bearing angles of four acoustic sources are set to $-33^\circ$, $-27^\circ$, $-12^\circ$ and $-3^\circ$, respectively. The distances between acoustic sources and the reference point are all set to $3$m.

\begin{figure}[htbp]
\centering
\includegraphics[width=2in]{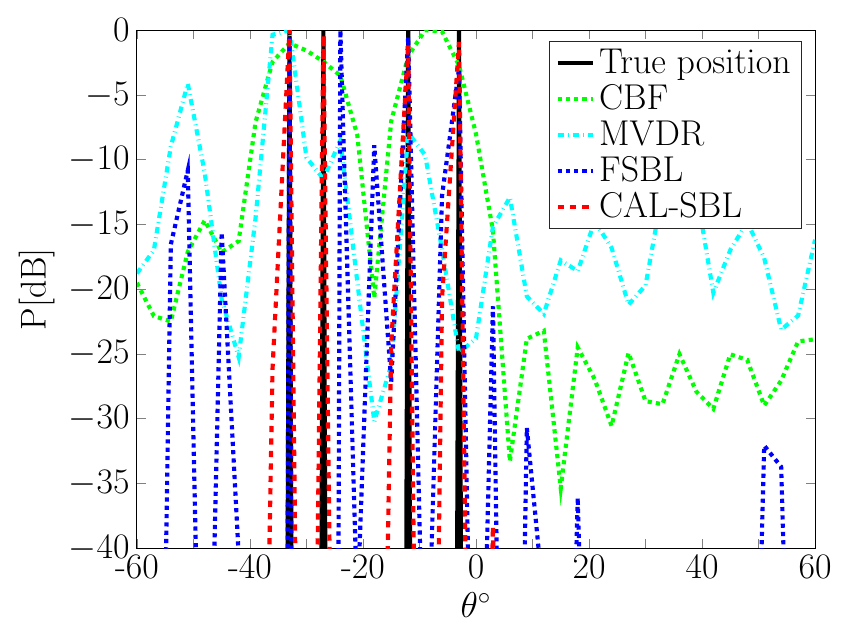}
\caption{Resolution performance for different methods and the black lines denote true bearing angle.}
\label{fig_doa_res}
\end{figure}

The clean synthetic microphone array data is generated using `RIR-generator' software with predefined parameters then white Gaussian noise is added to clean data. The SNR is set to $10$dB. 
The generated time-domain array data is converted to the frequency-domain array data using the short time Fourier transform (STFT). The length of each frame is set to $1024$ points and the length of increment is set to $256$ points, i.e., the overlap between frames is $75\%$. The total point number of fast Fourier transform (FFT) is set to $1024$. As all acoustic sources are $1$kHz pure sinusoidal acoustic source, we use the frequency bin contained $1$kHz to estimate acoustic DOAs. To build the dictionary $\bm{A}$, the target plane is separated into grids uniformly with an interval $1^\circ$. Then, acoustic DOA of each source is estimated using different algorithms, i.e., SRP-PHAT refers to steering response power phase transform (SRP-PHAT) based acoustic DOA algorithm proposed in \cite{Maximo2017}, MVDR refers to minimum variance distortionless response (MVDR) based algorithm proposed in \cite{An2020}, FSBL refers to SBL based algorithms proposed in \cite{Xenaki_2018} and CAL-SAVE refers to the proposed algorithm. Figure \ref{fig_doa_res} illustrates the spatial spectrum using different algorithms. 	

It can be seen from Figure \ref{fig_doa_res} that SRP-PHAT and MVDR methods fail to estimate DOA of all acoustic sources. FSBL algorithm estimates the two sources at $-12^\circ$ and $-3^\circ$ accurately, but fails to estimate the two sources at $-33^\circ$ and $-27^\circ$. However, the proposed algorithm estimates all the acoustic DOAs accurately, resulting in a higher resolution performance than state-of-the-art acoustic DOA methods. 

To further verify the estimation performance of the proposed algorithm for acoustic DOA estimation, we conduct experiments as follow. In this experiment, the room dimensions, RT60, reflection order and sound speed are exactly the same as in the previous experiment. An uniform linear microphone array is used to receive acoustic array signals. The inner space between adjacent microphones is set to $0.025$m. The center position of microphone array is set to $(5,2,1)$m. The number of acoustic sources is set to $K$ and all $K$ sources are located in the target plane, i.e., the plane in front of the uniform linear array, and we assume all sources located in the far-field of the microphone array. The distance between each source and the reference point is set to $3$m. The frequency values of all $K$ sources are all set to $1$kHz and initial phases of all sources are set randomly. The synthetic microphone array data can be generated using `RIR-generator' then white Gaussian noise is added. In this experiment, the acoustic DOA estimation performance are tested versus different numbers of microphones $M$, different numbers of sources $K$ and different SNRs, respectively.

To build the dictionary $\bm{A}$, the target plane is separated uniformly in the range from $-60^\circ$ to $60^\circ$ with an interval of $1^\circ$, i.e., the number of grids is $121$. After estimated acoustic DOA of all sources using different methods, the root mean square error (RMSE) is used to measure the estimation performance, which is defined as
\begin{equation}
e_{\rm RMSE}=\sqrt{\frac{1}{KN_{MC}}\sum_{n=1}^{N_{MC}}\sum_{k=1}^{K}\big(\hat{\theta}_{k}^n-\theta_{k}^n\big)^2},
\end{equation}
where $\theta_{k}^n$ denotes the truth DOA of the $k$'th acoustic source at the $n$'th MC experiment, whereas $\hat{\theta}_{k}^n$ is the estimation of $\theta_{k}^n$, and $N_{MC}$ is the total number of MC experiments. 

\begin{figure}
\begin{subfigure}[t]{0.32\textwidth}
	\centering
	\includegraphics[width=1.55in]{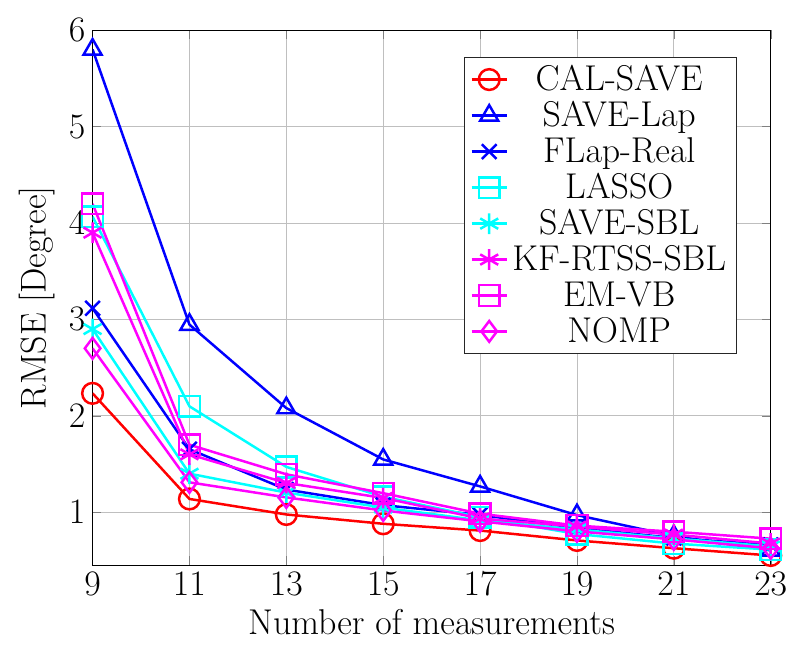}
	\caption{}
	\label{fig_SVDic_RMSE_vsMeasure_GaussianSig}
\end{subfigure}
\begin{subfigure}[t]{0.32\textwidth}
	\centering
	\includegraphics[width=1.6in]{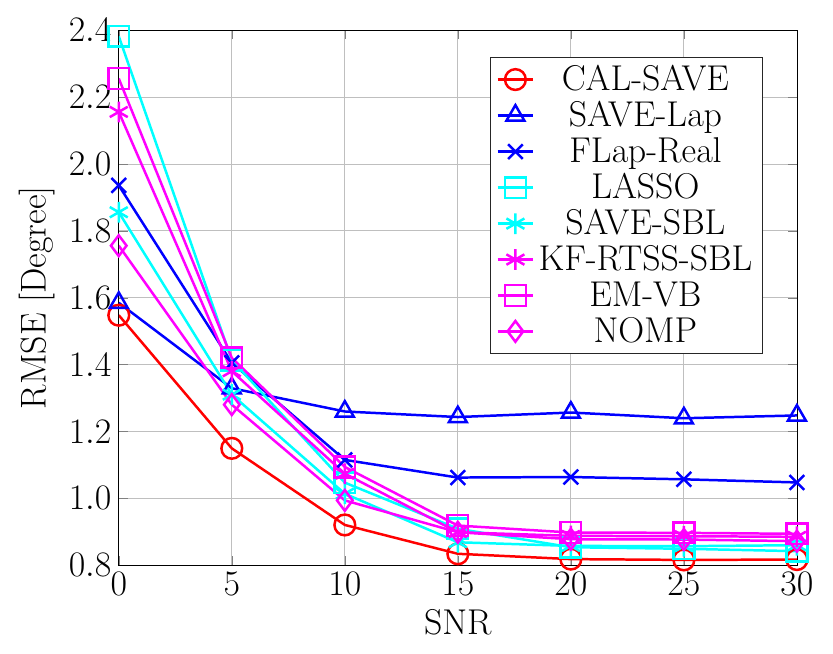}
	\caption{}
	\label{fig_SVDic_RMSE_vsSNR_LapSig}
\end{subfigure}
\begin{subfigure}[t]{0.32\textwidth}
	\centering
	\includegraphics[width=1.551in]{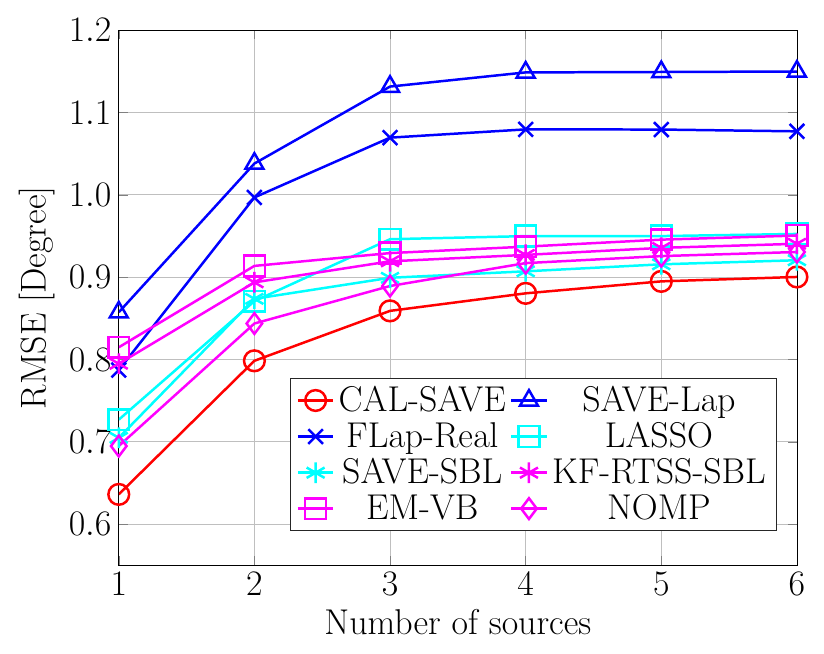}
	\caption{}
	\label{fig_SVDic_RMSE_vsNumSource_GaussianSig}
\end{subfigure}
\caption{Acoustic DOA estimation results. (a) RMSE of DOA estimation versus different length of measurement vector, (b) RMSE of DOA estimation versus different SNRs, (c) RMSE of DOA estimation versus different number of sources.}
\end{figure}

In the first experiment, the localization performance is tested versus different numbers of microphones in the range from $9$ to $23$ with an interval of $2$. The number of sources is set to $5$. To meet the far-field assumption, the distance between each source and the center point of microphone array is set to $3$m, and the bearing angles of all sources are randomly generated in the range from $-60^\circ$ to $60^\circ$. The SNR is set to $10$ dB. The number of MC experiments is set to $200$. The estimation accuracy results of all methods are shown in Figure \ref{fig_SVDic_RMSE_vsMeasure_GaussianSig}. 
It can be seen from Figure \ref{fig_SVDic_RMSE_vsMeasure_GaussianSig} that a large number of microphones leads to more accurate estimation performance. Note that `CAL-SAVE' outperforms the other methods when the number of microphones is smaller than $15$.

In the second experiment, the performance is tested versus different SNRs from $0$ to $30$ dB with an interval of $5$ dB. The number of sources is set to $5$. The number of microphones is set to $15$. The accuracy performance of different methods is illustrated in Figure \ref{fig_SVDic_RMSE_vsSNR_LapSig}. As can been from Figure \ref{fig_SVDic_RMSE_vsSNR_LapSig}, the localization accuracy increases as SNR increases in the range from $0$ to $20$ dB. When the SNR is bigger than $20$ dB, the localization performance becomes stable. Similar to the first experiment, the proposed `CAL-SAVE' method outperforms the state-of-the-art methods in most of cases.

In the third experiment, the performance is tested versus different number of sources from $1$ to $6$ with an interval of $1$. The SNR is set to $10$ dB. The number of microphones is set to $15$. The estimation accuracy results are shown in Figure \ref{fig_SVDic_RMSE_vsNumSource_GaussianSig}. As can be seen from this figure, the proposed method performs the best with regard to different number of acoustic sources.

\section{Conclusion}\label{sec:conclusion}
In this paper, the complex-value sparse signal recovery problem in considered. Motivated by the self-regularizing nature of the Bayesian framework and the structure of the adaptive LASSO, we build a hierarchical Bayesian model using adaptive Laplace priors to pursue complex-value sparse signals. Moreover, the space alternating strategy is integrated into the proposed algorithm to avoid matrix inverse operation. By exploiting the sparsity nature of complex-value signals using adaptive Laplace priors, the proposed method achieves accurate recovery performance. In the experiment part, the performance of the proposed method is verified using complex Gaussian random dictionaries and acoustic DOA estimation, respectively. The experimental results show that the proposed hierarchical Bayesian framework with adaptive Laplace priors improves the recovery performance for all of the types of signals studied and improve the acoustic DOA estimation performance. Besides, the proposed hierarchical framework and algorithm are easy to implement and apply in complex signal recovery application.

% or

\section*{Acknowledge}
This work is supported by the Fundamental Research Funds for the Central Universities (No.2022MS077).

\appendix
%	\section{Details of derivation}\label{Appendix_A}
%	Considering the first two stages of hierarchy, the marginal distribution of $\bm{g}$ is
%	\begin{eqnarray}
%	p(\bm{g}|\bm{\gamma})
%	=  \int\prod\limits_{i=1}^N\mathcal{CN}(g_i;0,\lambda_i)\Gamma(\lambda_i;\frac{3}{2},\frac{\gamma_i}{4}){\rm d}\bm{\lambda},\\
%	=  \prod\limits_{i=1}^N \frac{2}{\pi^{\frac{3}{2}}}\bigg(\frac{\gamma_i}{4}\bigg)^{\frac{3}{2}}\int\lambda_i^{-\frac{1}{2}}e^{-\frac{\gamma_i}{4}\lambda_i-\frac{g_i^2}{\lambda_i}}{\rm d}\lambda_i.
%	\end{eqnarray}
%	Using the integration function
%	m \begin{equation}
%	\int_0^{\infty}x^{v-1}e^{-\beta x^p-\gamma x^{-p}}{\rm d}x=\frac{2}{p}\left(\frac{\gamma}{\beta}\right)^{\frac{v}{2p}}{\rm K}_{\frac{v}{p}}\left(2\sqrt{\beta\gamma}\right),
%	\end{equation}
%	we obtain
%	\begin{eqnarray}
%	v=\frac{1}{2}, p=1, \beta=\frac{\gamma_i}{4}, \gamma=  g_i^2.
%	\end{eqnarray}
%	Therefore, we obtain
%	\begin{eqnarray}
%	p(\bm{g}|\bm{\gamma})=  \prod\limits_{i=1}^N \frac{2}{\pi^{\frac{3}{2}}}\bigg(\frac{\gamma_i}{4}\bigg)^{\frac{3}{2}}2\bigg(\frac{4g_i^2}{\gamma_i}\bigg)^{\frac{1}{4}}{\rm K}_{\frac{1}{2}}\bigg(2\sqrt{\frac{\gamma_i g_i^2}{4}}\bigg),\\
%	=  \frac{\prod\limits_{i=1}^N\gamma_i}{(2\pi)^{N}}e^{-\sum\limits_{i=1}^N\sqrt{\gamma_i}|g_i|}
%	\end{eqnarray}
\section{Derivation of hidden parameters for the SMV case}\label{Appendix_B}
\paragraph{Update of $q_{\lambda}(\bm{\lambda})$}
The logarithmic approximate posterior of the variable $\lambda_i$ is
\begin{equation}
\ln q(\lambda_i)
=-\frac{1}{2}\ln\lambda_i-\frac{1}{4}{\rm E}_{q(\gamma_i)}(\gamma_i)\lambda_i-\lambda_i^{-1}{\rm E}_{q(\bm{g})}(g_i^{*}g_i).
\end{equation}
Note that the Gamma distribution is not the conjugate prior of the complex Gaussian distribution with a known mean. Thus, we use a generalized inverse Gaussian distribution to represent the approximate posterior distribution of $q_{\lambda}(\bm{\lambda})$. The parameters of this generalized inverse Gaussian distribution are $p_{\lambda}=\frac{1}{2}$, $a=\frac{1}{2}{\rm E}_{q(\gamma_i)}(\gamma_i)$ and $ b=2{\rm E}_{q(\bm{g})}(g_i^{*}g_i)$, where ${\rm E}_{q(\bm{g})}(g_i^{*}g_i)=\mu_{g_i}^{*}\mu_{g_i}+\Sigma_{g_{ii}}$, $\mu_{g_i}$ and $\Sigma_{g_{ii}}$ are the $i$th element of the vector $\bm{\mu_g}$ and the $i$th diagonal element of the matrix $\bm{\Sigma}_g$. 
Therefore, the means of $\lambda_i$ and $\lambda_i^{-1}$ are
\begin{align}
{\rm E}_{q(\lambda_i)}(\lambda_i)&=  \frac{\sqrt{b}{\rm K}_{\frac{3}{2}}(\sqrt{ab})}{\sqrt{a}{\rm K}_{\frac{1}{2}}(\sqrt{ab})}
=\frac{\sqrt{b}}{\sqrt{a}}+\frac{1}{a},\label{update_lambda}\\
{\rm E}_{q(\lambda_i)}(\lambda_i^{-1})&= \frac{\sqrt{a}{\rm K}_{\frac{3}{2}}(\sqrt{ab})}{\sqrt{b}{\rm K}_{\frac{1}{2}}(\sqrt{ab})}-\frac{2p}{b}
=\frac{\sqrt{a}}{\sqrt{b}}.\label{update_lambdaInv}
\end{align}

\paragraph{Update of $q(\bm{\gamma})$}
The approximate posterior of $\gamma_i$ is
\begin{equation}
\ln q(\gamma_i)
=\bigg(c+\frac{1}{2}\bigg)\ln\gamma_i-\Big(\frac{1}{4}{\rm E}_{q(\lambda_i)}(\lambda_i)+d\Big)\gamma_i,
\end{equation}
which indicate that $\gamma_i$ follows a Gamma distribution with the parameters
$\alpha_{\gamma}=\frac{3}{2}+c$ and
$\beta_{\gamma}=\frac{1}{4}{\rm E}_{q(\lambda_i)}(\lambda_i)+d$.
Therefore, the mean of $\gamma_i$ is
\begin{eqnarray}
{\rm E}_{q(\gamma_i)}=\frac{\alpha_{\gamma}}{\beta_{\gamma}}.\label{update_gamma}
\end{eqnarray}

\paragraph{Update of $q(\rho)$}
The approximate posterior of $\rho$ is
\begin{equation}
\ln q(\rho)= (a+M-1)\ln\rho-\Big({\rm E}_{q(\bm{g})}\big(\left\|\bm{x}-\bm{A}\bm{g}\right\|^2\big)+b\Big)\rho,\nonumber
\end{equation}
which indicates that $\rho$ follows a gamma distribution with the parameters 
$\alpha_{\rho}=M+a$ and $ \beta_{\rho}={\rm E}_{q(\bm{g})}\big(\left\|\bm{x}-\bm{A}\bm{g}\right\|^2\big)+b$,
where
\begin{eqnarray}
{\rm E}_{q(\bm{g})}\big(\left\|\bm{x}-\bm{A}\bm{g}\right\|^2\big)=\left\|\bm{x}-\bm{A}{\rm E}_{q(\bm{g})}(\bm{g})\right\|^2+{\rm tr}(\bm{\Sigma}_g\bm{A}^{\rm H}\bm{A}).\nonumber
\end{eqnarray}
Thus, the mean of $\rho$ is
\begin{eqnarray}
{\rm E}_{q(\rho)}=\frac{\alpha_{\rho}}{\beta_{\rho}}.\label{update_rho}
\end{eqnarray}

\section{Derivation of hidden parameters for the MMV case}\label{Appendix_MMV}
\paragraph{Update of $q(\bm{\lambda})$}
Similar to the SMV case, the approximate posterior of $\lambda_i$ can be written as
\begin{align*}
\ln q(\lambda_i)&= {\rm E}_{q(\bm{\theta}\backslash \lambda_i)}\big(p(\bm{X},\bm{\theta})\big),\\
&=-\frac{1}{2}\ln\lambda_i-\frac{1}{4}{\rm E}_{q(\gamma_i)}(\gamma_i)\lambda_i-\lambda_i^{-1}{\rm E}_{q(g_i)}\Big({\rm{tr}}(\bm{g}_{i\cdot}^{\rm H}\bm{g}_{i\cdot})\Big)+c_{\lambda},
\end{align*}
which indicates that $\lambda_i$ follows an inverse Gaussian distribution with the parameters 
$p=\frac{1}{2}$, $a=\frac{1}{2}{\rm E}_{q(\gamma_i)}(\gamma_i)$ and
$b=2{\rm E}_{q(g_i)}\Big({\rm{tr}}(\bm{g}_{i\cdot}^{\rm H}\bm{g}_{i\cdot})\Big)$,
where ${\rm E}_{q(g_i)}\Big({\rm{tr}}(\bm{g}_{i\cdot}^{\rm H}\bm{g}_{i\cdot})\Big)=\bm{\mu}_i^{\rm H}\bm{\mu}_i+L\sigma_i$.
Thus, we obtain
\begin{align}
{\rm E}_{q(\lambda_i)}(\lambda_i)&=\frac{\sqrt{b}}{\sqrt{a}}+\frac{1}{a}.\label{update_lambda_MMV}\\
{\rm E}_{q(\lambda_i)}\big(\lambda_i^{-1}\big)&=\frac{\sqrt{a}}{\sqrt{b}}.\label{update_lambdaInv_MMV}
\end{align}

\paragraph{Update of $q(\bm{\gamma})$}
The (\ref{evi_MMV}) leads to a Gamma distribution for the approximate distribution of $\gamma_i$, that is
\begin{align*}
\ln q(\gamma_i)&= {\rm E}_{q(\bm{\theta}\backslash \gamma_i)}\big(p(\bm{X},\bm{\theta})\big),\\
&= \bigg(L+c-\frac{1}{2}\bigg)\ln\gamma_i-\bigg(\frac{1}{4}{\rm E}_{q(\lambda_i)}(\lambda_i)+d\bigg)\gamma_i+c_{\gamma},
\end{align*}
with the parameters
$\alpha_{\gamma}=L+c+\frac{1}{2}$ and $\beta_{\gamma}=\frac{1}{4}{\rm E}_{q(\lambda_i)}(\lambda_i)+d$.
Thus, the mean of $\gamma_i$ for the MMV case is  
\begin{eqnarray}
{\rm E}_{q(\gamma_i)}(\gamma_i)=\frac{\alpha_{\gamma}}{\beta_{\gamma}}.\label{update_gamma_MMV}
\end{eqnarray}

\paragraph{Update of $q(\rho)$}
Similarly, the approximate distribution of $\rho$ is
\begin{align*}
\ln q(\rho)&={\rm E}_{q(\bm{\theta}\backslash \rho)}\big(p(\bm{X},\bm{\theta})\big),\nonumber\\
&=(LM+a-1)\ln\rho-\Bigg({\rm E}_{q(\bm{G})}\bigg(\left\|\bm{X}-\bm{A}\bm{G}\right\|_f^2\bigg)+b\Bigg)\rho,\nonumber
\end{align*}
which indicates that $\rho$ follows a Gamma distribution with the parameters as follows:
\begin{equation}
\alpha_{\rho}=LM+a,
\beta_{\rho}={\rm E}_{q(\bm{G})}\Big(\left\|\bm{X}-\bm{A}\bm{G}\right\|_f^2\Big)+b,\nonumber
\end{equation}
where 
\begin{equation}
{\rm E}_{q(\bm{G})}(\left\|\bm{X}-\bm{A}\bm{G}\right\|_f^2)= \left\|\bm{X}-\bm{A}{\rm E}_{q(\bm{G})}(\bm{G})\right\|^2+{\rm tr}(\bm{\Sigma}\bm{A}^{\rm H}\bm{A}).\nonumber
\end{equation}
Thus, we have
\begin{equation}
{\rm E}_{q(\rho)}(\rho)=\frac{\alpha_{\rho}}{\beta_{\rho}}.\label{update_rho_MMV}
\end{equation}

\bibliographystyle{unsrt}
\bibliography{main.bib}

\end{document}